\documentclass[12pt]{article}

\usepackage{hyperref}

\usepackage{appendix}
\usepackage{amsmath, amssymb, mathrsfs}
\usepackage{tikz, tcolorbox, color}
\usetikzlibrary{decorations.markings, arrows.meta, calc}
\tcbuselibrary{skins}
\usepackage {diagbox}
\usepackage{ulem}
\usepackage{physics}
\usepackage{fancybox}
\usepackage{fancyhdr}
\usepackage{bm}

\usepackage{epsf}
\usepackage{amsmath,amssymb}
\usepackage{latexsym}
\usepackage{graphicx,color}
\usepackage{wrapfig}
\usepackage{latexsym}
\usepackage{graphics}

\usepackage{color}
\usepackage{delarray,color} 

\usepackage[T1]{fontenc}
\usepackage{lmodern}
\usepackage[utf8]{inputenc}

\usepackage{mathtools}

\newcommand {\beq}{\begin{align}}
\newcommand {\eeq}{\end{align}}

\newcommand{\be}{\begin{equation}}
\newcommand{\ba}{\begin{align}}
\newcommand{\ea}{\end{align}}
\newcommand{\ee}{\end{equation}}

\newcommand{\beqa}{\begin{align}}
\newcommand{\eeqa}{\end{align}}

\newcommand{\unit}{\hbox to 3.8pt{\hskip1.3pt \vrule height 7.4pt
    width .4pt \hskip.7pt \vrule height 7.85pt width .4pt \kern-2.4pt
    \hrulefill \kern-3pt \raise 3.7pt\hbox{\char'40}}}

\def\matt[#1,#2,#3,#4]{\left(%
\begin{array}{cc} #1 & #2 \\ #3 & #4 \end{array} \right)}

\newcommand{\loc}{\mathrm{loc}}
\newcommand{\atm}{\mathrm{atm}}

\usetikzlibrary{arrows.meta,positioning,calc,decorations.pathmorphing,patterns}

\usepackage{tabularx}

\catcode`\@=11
\@addtoreset{equation}{section}

\catcode`@=12
\relax

\begin{document}

\begin{titlepage}

\setcounter{page}{0}

\renewcommand{\thefootnote}{\fnsymbol{footnote}}

\begin{flushright}

YITP-26-62

\end{flushright}

\vskip 1.35cm

\begin{center}
{\Large \bf 
Finite \(N\) Black Holes through the Brick Wall
}

\vskip 1.2cm 

{\normalsize
Seiji Terashima${}^{a}$~\footnote{terasima(at)yukawa.kyoto-u.ac.jp}
}

\vskip 0.8cm


${}^{a}${\it
Center for Gravitational Physics and Quantum Information, \\
Yukawa Institute for Theoretical Physics, Kyoto University, Kyoto 606-8502, Japan
}

\end{center}

\vspace{12mm}

\begin{abstract}
In AdS/CFT at finite \(N\), black holes are described by ordinary quantum systems with finite entropy and discrete spectra, while semiclassical bulk effective field theory treats the near-horizon region as an ideal continuum.  
We reconsider the brick-wall model as a simple effective description of
the finite \(N\) departure from the semiclassical near-horizon continuum in
AdS/CFT.
A Dirichlet wall is reflecting; however, once the regulated near-wall modes are replaced by interacting stretched-horizon degrees of freedom, the exterior simple sector can look absorptive and black-hole-like.  If these degrees of freedom are thermally populated while the exterior incoming modes are in their vacuum, they radiate as ordinary hot matter, and this gives Hawking radiation.
Furthermore, we show that finite \(N\) prevents the near-horizon sector from being an ideal continuum absorber; the brick-wall model is a simple effective representation of this limitation, where the failure of perfect absorption appears as residual reflection 
or echoes.  
Such return effects appear on the echo time scale
\(\kappa^{-1}\log S_{\rm BH}\) and are not expected to be
suppressed as \(e^{-S_{\rm BH}}\), although their detailed amplitude is
model-dependent.
We also discuss brick-wall-like 
models in which the effective inner boundary arises dynamically from a regular matter core rather than from an imposed boundary condition.
\end{abstract}

\end{titlepage}

\newpage

\tableofcontents
\vskip 1.2cm

\setcounter{footnote}{0}

\section{Introduction}

In AdS/CFT \cite{Maldacena:1997re} \cite{Gubser:1998bc, Witten:1998qj}, a finite $N$ black hole is described by an ordinary quantum
system with finite entropy and a discrete spectrum \cite{Maldacena:2001kr,Barbon:2003aq,Kleban:2004rx, Festuccia:2006sa, Kabat:2014kfa}.
On the other hand,
semiclassical bulk effective field theory (EFT) treats the near-horizon region as
an ideal continuum.\footnote{
By bulk EFT we mean the low-energy effective description of the bulk,
including semiclassical gravity and matter fields.
}  
In particular, the usual black hole EFT description
allows an incoming perturbation to disappear into the horizon as if the
horizon were a perfect continuum absorber.  This is an excellent
coarse-grained approximation in the semiclassical regime, but it need not
be 
a finite $N$ statement.\footnote{
Here, finite $N$ (though very large) corresponds to finite $G_N$ (though very small) in the bulk gravity theory; we are concerned with effects at finite $G_N$, not with the $1/N$ expansion around $N=\infty$.
}

The brick-wall model of 't Hooft \cite{tHooft:1984kcu} was originally introduced as a simple regulator of the divergent density of quantum-field modes near a black-hole horizon.  In this model, the exterior fields are quantized outside a timelike surface placed a small proper distance outside the horizon, and a boundary condition is imposed on this surface.  With a cutoff of order the Planck length, the resulting thermal atmosphere gives an entropy of the order of the Bekenstein--Hawking entropy \cite{Bekenstein:1973ur, Hawking:1975vcx}.
In this paper we reconsider the brick-wall model of 't Hooft
\cite{tHooft:1984kcu} from this finite $N$ point of view.\footnote{
There were some important results for the large $N$ limit of the brick-wall model, for example \cite{Susskind:1993if, Susskind:1994sm, Demers:1995dq, Solodukhin:1996xv, Mukohyama:1998rf}. For recent discussions on the brick-wall model, see \cite{Krishnan:2023jqn, Burman:2023kko,  Jeong:2024jjn, Jeong:2025jyx, Begines:2026fnx}.
}
We do not
claim that a black hole has a literal hard Dirichlet surface.  Rather, the
brick wall is used as a simple effective regulator for the failure of the
continuum near-horizon bulk EFT to provide finite state counting, as discussed in \cite{Iizuka:2013kma, Terashima:2021klf, Sugishita:2023wjm}.  It
replaces the infinite density of near-horizon modes by a large, but finite,
stretched-horizon reservoir.

There is an immediate concern with this viewpoint.  A Dirichlet wall by
itself gives a reflecting boundary condition.  In the free-field
approximation it describes a cavity, not an absorber, and it does not by
itself produce Hawking radiation \cite{Hawking:1975vcx}.  However, in the
present interpretation the wall is not a free mirror with no additional
dynamics.  The regulated near-wall modes are understood as the EFT modes
near the cutoff surface whose dynamics is modified by order-one
gravitational interactions, or equivalently as interacting degrees of
freedom associated with a stretched horizon.  An incoming simple excitation
can then be converted into complicated near-wall excitations, so that the
exterior simple sector can look absorptive and black-hole-like even though
the full microscopic evolution remains unitary.
The associated return is
therefore an ordinary response of the regulated near-wall dynamics, not a
Heisenberg-time recurrence arising from exponentially small overlaps of
black-hole microstates; there is no general reason for it to be suppressed
as \(e^{-S_{\rm BH}}\).

The basic point is the near-horizon blueshift.  A mode with small
Killing energy as measured at infinity is not necessarily low energy
locally.  Near a Planckian stretched horizon its local energy can be of
order the cutoff scale.  Therefore gravitational and other interactions in
the stretched-horizon layer need not be weak.  This is the mechanism, in
the brick-wall language, by which a reflecting regulator can nevertheless
give an effectively black hole response to exterior probes, as discussed in \cite{Terashima:2025tct}.

The same reservoir picture also gives a simple account of Hawking
radiation.  The Hartle--Hawking-like state is an equilibrium state in which
both the near-wall reservoir and the ordinary exterior modes are thermal at
the Hawking temperature \(T_H\), so there is no net luminosity.  The
Unruh-like state \cite{Unruh:1976db} is different: the near-wall reservoir is thermally, or
typically, populated, while the ordinary exterior incoming modes are in
their vacuum.  In asymptotically flat spacetime this is just a hot object
radiating into empty space.  After the usual greybody filtering, this
gives the Hawking flux.  Thus Hawking radiation is not derived from a free
reflecting wall; it is interpreted as black-body emission from an
interacting stretched-horizon reservoir.

Building on these earlier results, the present work focuses on the
finite-\(N\) relation between ordinary CFT evolution and Kruskal-like
evolution, and on the thermodynamic consistency of near-extremal
throat excitations.

This paper makes three main claims.

First, as already indicated above, we show that an interacting brick-wall model can be used as a consistent effective toy model of the near-horizon region.\footnote{
The membrane paradigm \cite{Price:1986yy} may look similar to the brick-wall picture, since both introduce a stretched surface just outside the horizon. However, their roles are different. The usual membrane paradigm is essentially a dissipative rewriting of the ingoing horizon boundary condition, and therefore keeps the standard black-hole retarded response. In contrast, the brick-wall picture used here treats the stretched surface as an effective finite $N$ replacement of the near-horizon continuum; perfect absorption is then not imposed as an exact boundary condition.
}

Second, we argue that finite \(N\) generically prevents the near-horizon region from being a perfect continuum absorber.  This conclusion is not tied to the brick-wall model; the brick wall is only the simplest effective way to display it.  In this description, the failure of perfect absorption appears as residual reflection from the regulated inner boundary, leading to a late-time return, or echo, of an initially absorbed perturbation.  We argue that this return is not a feature of the brick-wall model but a consequence of finite \(N\): combining the finite \(N\) state counting with the no-go result of Section~\ref{sec:finiteN-reflection}, that the smooth Kruskal evolution through the horizon is not realized by the CFT Hamiltonian, the equivalence principle must fail at a stretched horizon a Planckian distance outside the horizon.  The return then appears on the echo time scale \(\kappa^{-1}\log S_{\rm BH}\), and we argue that it is an \(O(N^0)\) effect, not an \(e^{-S_{\rm BH}}\) one.

A related 
finite \(N\) issue\footnote{
A possibly related finite \(N\) gap from the \(1/N\) expansion was discussed in \cite{Terashima:2025shl, Sugishita:2022ldv, Sugishita:2023wjm, Sugishita:2024lee} in the context of bulk reconstruction for subregions \cite{Terashima:2020uqu, Terashima:2017gmc, Terashima:2021klf}. 
}
concerns the status of different time evolutions in the bulk description.  In semiclassical bulk EFT, different slicings of the eternal black hole, including Kruskal-like slicings through the smooth horizon, are normally regarded as diffeomorphic descriptions of the same physics.  In the finite \(N\) CFT, however, time evolution is specified by a Hamiltonian.  Therefore the ordinary thermofield-double (TFD) evolution generated by \(H_L+H_R\), or by \(H_R-H_L\) depending on conventions, need not be identical as an exact finite-\(N\) Hamiltonian evolution to a Hamiltonian \(H_{\rm Kr}\) realizing a Kruskal-like smooth-horizon evolution, as discussed for the Rindler case with a lattice cutoff \cite{Chikazawa:2026iro}.
This point is logically separate from the brick-wall model, but it supports the same lesson: the strict semiclassical identification of continuum near-horizon evolution need not survive unchanged at finite \(N\).
This concerns a failure of the exact microscopic equivalence suggested by the semiclassical bulk description.

Third, the brick-wall-like cutoff need not always be imposed by hand.  In
suitable matter--gravity systems the black hole-like exterior can end on a
regular matter core rather than on an artificial boundary.  Gravitating
monopoles \cite{Breitenlohner:1991aa, Lee:1991vy, Lue:1999zp} provide a concrete example: the exterior can be close to a
charged black hole geometry down to a core which replaces the would-be
horizon or singular region.  The regularity 
at the
core then plays the role of an effective inner boundary condition.  This is
not a usual fuzzball construction, since one does not obtain an
exponentially large family of distinct classical geometries with the same
asymptotic charges.  It is, however, fuzzball-like in the weaker sense
relevant here: the would-be horizon is replaced by a cap, the exterior
remains black hole-like, and the brick-wall entropy is interpreted as the
entropy of semiclassical excitations supported in the redshifted region
outside the cap.


The viewpoint here is consistent with the central dogma in its minimal 
form: from the outside, a black hole is an ordinary quantum system with 
$S_{\rm BH}$ degrees of freedom and unitary evolution. What we do not assume is 
the stronger expectation that the exterior effective description through a 
smooth horizon remains accurate, with deviations suppressed only as 
$e^{-S_{\rm BH}}$, up to the Page or Heisenberg time. The statements made here 
instead follow from properties that hold by definition in a finite $N$ 
holographic CFT: in a fixed energy band the Hilbert space is finite-dimensional, and on a compact boundary the spectrum is discrete. A finite, 
discrete system cannot reproduce an exact continuum absorber at arbitrarily 
fine resolution, and the brick wall is used only to make this kinematical 
fact visible.

The organization is as follows.  Section~\ref{sec:finiteN-viewpoint}
explains why the brick wall is used as a finite $N$ effective regulator
for the near-horizon reservoir.  Section~\ref{sec:state-counting}
discusses the interacting brick-wall picture, the area-law state counting,
the effective absorptive response, and the Unruh-like state needed for
Hawking flux.  Section~\ref{sec:response-echoes} translates the same
physics into retarded Green functions, quasinormal response, and echoes.
Section~\ref{sec:finiteN-reflection} states the finite $N$ lesson for
thermal correlators and for the relation between ordinary TFD evolution and
Kruskal-like evolution.  Finally, Section~\ref{sec:near-extremal}
discusses near-extremal throats and explains how regular matter cores, such
as gravitating monopoles, can realize a brick-wall-like or fuzzball-like cap
without imposing a boundary condition by hand.

\section{A brief review of the brick-wall model and its finite $N$ interpretation}
\label{sec:finiteN-viewpoint}

Let us first recall the minimal idea of the brick-wall model \cite{tHooft:1984kcu}.  In the
original treatment, one considers quantum fields outside a static black
hole and introduces a cutoff surface at a small proper distance from the
horizon.  For example, for a Schwarzschild-like metric
\begin{align}
  ds^2=-f(r)dt^2+\frac{dr^2}{f(r)}+r^2 d\Omega^2 ,
  \qquad f(r_h)=0 ,
\end{align}
one replaces the horizon by a surface
\begin{align}
  r=r_h+h ,
\end{align}
or equivalently by a surface at fixed proper distance
\begin{align}
  \rho \sim
  \int_{r_h}^{r_h+h} \frac{dr}{\sqrt{f(r)}} .
\end{align}
A simple boundary condition, such as
\begin{align}
  \phi(r_h+h)=0 ,
\end{align}
is then imposed on the field.\footnote{
't Hooft \cite{tHooft:1984kcu} regarded the brick wall as an artificial cutoff because a literal
wall violates the equivalence principle at the horizon.  Here we use this
observation in the opposite direction: the equivalence principle is a
property of the semiclassical bulk EFT, and the question is whether it
remains an exact statement in finite $N$ quantum gravity.  The brick wall
is not assumed to be a fundamental surface, but is used as an effective
parametrization of a possible finite $N$ breakdown of the smooth-horizon
EFT.
} 
The near-horizon region has a large blueshift: a mode with fixed Killing
energy at infinity has a local energy
\begin{align}
  \omega_{\rm loc}={\omega\over \sqrt{f(r)}} ,
\end{align}
which becomes large near the horizon.  Equivalently, modes with finite
local momentum can have very small Killing energy at infinity.  As a
result, the density of modes outside the wall is large and diverges as the
wall is taken to the horizon.

In the free-field WKB estimate, the entropy of a thermal gas outside the
wall is dominated by the near-horizon modes and diverges as the wall is
taken to the horizon.  Requiring this entropy not to exceed the
Bekenstein--Hawking entropy fixes the proper distance of the wall to be of
order the Planck length.  With such a cutoff one obtains
\begin{align}
  S_{\rm wall}\sim {A\over G_N},
\end{align}
up to a numerical coefficient depending on the number of fields and on the
precise cutoff prescription.  
This is the usual sense in which the brick-wall model gives an entropy of
the same order as the Bekenstein--Hawking entropy.


We next isolate the meaning of the brick wall in finite $N$ AdS/CFT,
where a holographic CFT is taken as the microscopic definition of quantum gravity.
This prevents a possible misunderstanding: the wall is not introduced as a
literal material surface, but as a simple way to encode the failure of the continuum near-horizon
bulk EFT state counting.

In \cite{Iizuka:2013kma}, a holographic interpretation of brick-wall physics was given.  
In the large-$N$ deconfined phase, the boundary spectrum becomes effectively continuous, and the
bulk probe free energy develops a near-horizon divergence.  At finite $N$, the boundary spectrum
is discrete, and the divergence is removed.  From the bulk viewpoint, the conclusion is that naive
bulk effective field theory overcounts the number of independent near-horizon states.

This is the sense in which we use the brick-wall model.  We do not claim that a finite $N$ literally
produces a Dirichlet boundary condition at a sharply defined proper distance.  The wall is a toy
regulator which imitates a qualitative finite $N$ feature: the infinite continuum of semiclassical
near-horizon modes must be replaced by a finite, though very large, set of microscopic degrees of
freedom.  

This viewpoint is close in spirit to the general lesson of the fuzzball conjecture
\cite{Mathur:2005zp, Mathur:2009hf}.  In a fuzzball
picture, a black hole microstate is not literally the eternal smooth horizon geometry continued
all the way inward; instead the near-horizon region is replaced by microstructure.  The present
paper does not use the detailed string-theoretic construction of fuzzballs, nor does it require that
the microstructure be a classical fuzzball geometry.\footnote{
For a discussion of stretched horizons in the fuzzball context
within AdS/CFT, based on coarse-graining of LLM bubbling geometries,
see \cite{Ogawa:2008kz}.
}  
The claim is weaker and more universal:
for a generic finite $N$ black hole, the smooth-horizon EFT should be replaced, at sufficiently
short proper distance from the horizon, by some finite microscopic structure.  The brick-wall model is a simple phenomenological model of this replacement.\footnote{
As shown in \cite{Susskind:1993if, Susskind:1994sm, Demers:1995dq}, the divergence of the entropy with the brick wall can be absorbed by the renormalization of $G_N$.
This is the consistency of the large $N$ expansion around $N=\infty$, which shows that this expansion makes sense. 
Here, we claim that 
the finite $N$ result is different from the large $N$ expansion.
The finite number of degrees of freedom in the holographic CFT implies
that the naive continuum bulk EFT cannot be extended beyond the brick-wall
cutoff as an exact description.
}


\section{Interacting brick wall, black hole thermodynamics, and Hawking radiation}
\label{sec:state-counting}

In this section, we reorganize the usual brick-wall discussion in the form needed for the rest of the paper.  The final claim is not merely that a brick wall can reproduce the area scaling of the entropy.  
Rather, the gravitational interaction near the brick wall should be included through order-one effective coefficients.  With this interpretation, the brick-wall model gives a self-consistent effective description of black-hole thermodynamics and Hawking emission.\footnote{
Most claims in this section were essentially obtained in \cite{Iizuka:2013kma, Terashima:2025tct}.
Here, we clarify such results for the consistent description of the black hole thermodynamics.
}


Consider a static black hole in $D$ spacetime dimensions,
\begin{equation}
  ds^{2}= -f(r)dt^{2}+\frac{dr^{2}}{f(r)}+r^{2}d\Omega_{D-2}^{2},
  \qquad
  f(r_h)=0,
  \qquad
  \kappa=\frac12 f'(r_h),
  \qquad
  T_H=\frac{\kappa}{2\pi} .
\end{equation}
The horizon area is
\begin{equation}
  A_h=\Omega_{D-2}r_h^{D-2} .
\end{equation}
Here \(\Omega_{D-2}=2\pi^{(D-1)/2}/\Gamma((D-1)/2)\) is the area of the unit \((D-2)\)-sphere, and \(\kappa\) is the surface gravity.
Near the horizon,
\begin{equation}
  f(r)\simeq 2\kappa(r-r_h).
\end{equation}
The proper distance from the horizon is
\begin{equation}
  \rho(r)=\int_{r_h}^{r}\frac{dr'}{\sqrt{f(r')}}
  \simeq \sqrt{\frac{2(r-r_h)}{\kappa}},
\end{equation}
and hence
\begin{equation}
  r-r_h\simeq \frac12\kappa\rho^2,
  \qquad
  f\simeq \kappa^2\rho^2.
\end{equation}
The brick wall or stretched horizon is placed at a fixed proper distance
\begin{equation}
  \rho=\epsilon,
\end{equation}
which we will take the Planck length, i.e. $\epsilon\sim l_{p}$,
where $r_\epsilon$ is the position of the brick wall.
Equivalently,
\begin{equation}
  r_\epsilon-r_h\simeq \frac12\kappa\epsilon^2.
\end{equation}
We here use $\epsilon$ as a proper-distance cutoff.  This is important because the leading powers in the entropy and energy are powers of the proper distance from the horizon.

The local temperature associated with a state at the Hawking temperature is
\begin{equation}
  T_\loc(\rho)=\frac{T_H}{\sqrt{-g_{tt}}}
  \simeq \frac{T_H}{\kappa\rho}
  =\frac{1}{2\pi\rho}.
  \label{eq:Tolman}
\end{equation}
Thus a stretched horizon placed at $\rho\sim l_{p}$ is locally at a Planckian temperature, although the red-shifted temperature measured at infinity is $T_H$.  A mode with asymptotic frequency $\omega\sim T_H$ has local frequency
\begin{equation}
  \omega_\loc(\rho)\simeq \frac{\omega}{\kappa\rho},
\end{equation}
so at the stretched horizon it is a cutoff-scale excitation.  In this sense, the modes are low energy only with respect to the asymptotic Killing time.  Locally, the same modes probe the Planckian physics.

\subsection{Free WKB state counting and what it does not prove}

First recall the standard free-field estimate.  In the Hartle--Hawking state, the exterior quantum fields are thermal at temperature $T_H$ with respect to the asymptotic time $t$.  
We will consider such a finite temperature state for the brick-wall model. 
Near the horizon, this may be evaluated locally using \eqref{eq:Tolman}.  For a massless conformal gas,
\begin{equation}
  e(\rho)=a_D (T_{\rm loc})^D,
  \qquad
  s(\rho)=\frac{D}{D-1}a_D (T_{\rm loc})^{D-1},
\end{equation}
where $a_D$ includes the effective number of species, $e$ is the energy density, $s$ is the entropy density and $T_{\rm loc}$ is the local temperature.  The proper volume element in the leading near-horizon region is
\begin{equation}
  dV_{\rm prop}\simeq A_h\,d\rho .
\end{equation}
The entropy of the thermal atmosphere $S_\atm$ is therefore
\begin{align}
  S_\atm
  &\simeq A_h\int_\epsilon d\rho\,s(\rho)
   \simeq A_h\frac{D}{D-1}a_D
   \int_\epsilon \frac{d\rho}{(2\pi\rho)^{D-1}} \\
  &\simeq
  \frac{A_h}{\epsilon^{D-2}}
  \frac{D}{D-1}\frac{a_D}{(2\pi)^{D-1}(D-2)}+\text{subleading} .
  \label{eq:Satm}
\end{align}
The local energy diverges more strongly, but the Killing energy $E$ measured at infinity contains one red-shift factor,
\begin{equation}
  dE=\sqrt{-g_{tt}}\,e\,dV_{\rm prop}
  \simeq \kappa\rho\,e\,A_h d\rho .
\end{equation}
Hence, the energy of the thermal atmosphere $E^\atm$ is given by
\begin{align}
  E^\atm
  &\simeq A_h a_D\int_\epsilon d\rho\,
  \kappa\rho (2\pi\rho)^{-D} \\
  &\simeq
  \frac{A_h\kappa}{\epsilon^{D-2}}
  \frac{a_D}{(2\pi)^D(D-2)}+	\text{subleading}
  =\frac{D-1}{D}T_H S_\atm+	\text{subleading} .
  \label{eq:Einfty}
\end{align}
Thus, the free WKB brick-wall gas gives
\begin{equation}
  S_\atm=\frac{D}{D-1}\, \frac{1}{T_H} E^\atm
  \label{eq:free-gas-factor}
\end{equation}
for the leading conformal contribution,
as expected.

The same result follows directly from WKB mode counting.  For a fixed asymptotic frequency $\omega$, the radial momentum near the horizon obeys schematically
\begin{equation}
  p_\rho^2+k_\perp^2+m^2\simeq \frac{\omega^2}{\kappa^2\rho^2}.
\end{equation}
Therefore the allowed largest transverse momentum grows as
\begin{equation}
  k_\perp^{\rm max}(\rho)\sim \frac{\omega}{\kappa\rho} .
\end{equation}

Equivalently, one may estimate the number \(N(\omega)\) of
WKB modes with Killing energy below \(\omega\) by integrating the local
phase-space volume over the near-horizon region, schematically,
$
  N(\omega)\sim
  \int d^{D-1}x\,d^{D-1}p\,
  \Theta\!\left(\omega-H(x,p)\right)
$.
Thus, up to numerical factors,
\begin{align}
  N(\omega)
  \sim
  \int_{\epsilon}^{\rho_{\rm out}} d\rho\, A_h
  \int {dp_\rho\,d^{D-2}k_\perp\over (2\pi)^{D-1}}\,
  \Theta\!\left(
    {\omega^2\over \kappa^2\rho^2}
    -p_\rho^2-k_\perp^2-m^2
  \right).
\end{align}
Near the wall, the mass term is subleading, and the momentum integral gives
a factor of order \((\omega/\kappa\rho)^{D-1}\).  Hence
\begin{equation}
    N(\omega)
  \sim A_h\left({\omega\over\kappa}\right)^{D-1}
  \int_\epsilon^{\rho_{\rm out}}{d\rho\over \rho^{D-1}}
  \sim
  {A_h\over \epsilon^{D-2}}
  \left({\omega\over\kappa}\right)^{D-1}.
  \label{eq:Nomega}
\end{equation}
up to numerical and species-dependent factors.  Using
\begin{equation}
  F(\beta)=\frac{1}{\beta}\sum_n\log(1-e^{-\beta\omega_n})
  \simeq -\int_0^\infty d\omega\,\frac{N(\omega)}{e^{\beta\omega}-1},
\end{equation}
one obtains
\begin{equation}
  F(\beta)\propto -\frac{A_h}{\epsilon^{D-2}}\,\kappa^{-(D-1)}\beta^{-D} .
\end{equation}
The power $F\propto -\beta^{-D}$ is what gives \eqref{eq:free-gas-factor}.  Thus the factor $D/(D-1)$ is not a mysterious property of the wall.  It is the equation-of-state factor of a free, locally conformal, massless thermal gas.

This observation is important because it also tells us what is not universal.  If the wall is placed at a Planckian proper distance, then $T_\loc\sim 1/l_{p}$ and gravitational interactions in the stretched-horizon layer are expected to be order one.  The free WKB relation \eqref{eq:free-gas-factor} should therefore not be treated as a universal prediction of the model.  It is more appropriate to replace the free-gas coefficients by order-one constants,
\begin{equation}
  S_{wall}=c_S\frac{A_h}{l_{p}^{D-2}},
  \qquad
  E_{wall}=c_E T_H\frac{A_h}{l_{p}^{D-2}},
  \qquad
  c_S,c_E=O(1),
  \label{eq:interacting-SE}
\end{equation}
where $c_S$ and $c_E$ include the effects of strong gravitational interactions, the true UV degrees of freedom, the precise cutoff prescription, and any species dependence after the renormalization of $G_N$.  In the free gas approximation, one has a definite value for $c_S, c_E$, but in the interacting brick-wall model, we only assume $c_S,c_E=O(1)$.

The robust statement is therefore the scaling, not the numerical coefficient:
\begin{equation}
  S_{wall} \sim \frac{A_h}{G_N},
  \qquad
  E_{wall} \sim T_H\frac{A_h}{G_N}.
  \label{eq:robust-scaling}
\end{equation}
This is the scaling required for consistency with black hole thermodynamics.

For a Schwarzschild black hole,
\begin{equation}
  T_H S_{\rm BH}\sim M_{\rm ADM},
\end{equation}
Hence, the stretched-horizon reservoir carries an energy of the same order as the black hole mass and an entropy of the same order as the Bekenstein--Hawking entropy \cite{Bekenstein:1973ur, Hawking:1975vcx}.

This should not be interpreted as adding a second energy on top of the black hole mass.  The fixed-background thermal-atmosphere calculation sees a large near-horizon energy because it treats the background geometry as external.  In the gravitational interpretation, the stretched-horizon reservoir is an effective description of the microscopic degrees of freedom already responsible for the black hole entropy and mass.  Thus \eqref{eq:robust-scaling} is a consistency condition, not a double counting of the ADM energy.

It is important to distinguish the equation of state from the first-law
relation.  The free conformal-gas result
\begin{align}
  S_{\rm atm}={D\over D-1}\beta_H E^{\rm atm},
\end{align}
is not universal; it depends on the free, locally conformal equation of
state.  Once the stretched-horizon layer is Planckian and strongly
interacting, this finite relation can be modified by order-one coefficients.
This does not mean, however, that the thermodynamic first law
is lost.  For any ordinary statistical system in equilibrium at the fixed
background temperature \(T_H\), infinitesimal variations obey
\begin{align}
  dE = T_H\, dS .
\end{align}
The qualification is that in the present calculation the geometry is held
fixed, so \(T_H\) is treated as an external temperature.  Thus this should
be understood as the first law for small changes of the atmosphere or
stretched-horizon reservoir on a fixed black hole background, not as a
finite variation of the entire black hole solution.  In this sense the
interacting reservoir may have a different equation of state from the free
brick-wall gas while still satisfying the first-law relation automatically.

The divergent form of \eqref{eq:Satm}, in the $\epsilon \rightarrow 0$ limit,  may also be viewed in the standard way as the matter contribution to the renormalization of Newton's constant \cite{Susskind:1994sm, Demers:1995dq},
\begin{equation}
  \frac{A_h}{4G_N^{\rm bare}}+S_\atm
  =\frac{A_h}{4G_N^{\rm ren}}+S_{\rm finite}.
\end{equation}
The present interpretation is slightly different in emphasis.  The
renormalization of \(G_N\) demonstrates the consistency of the semiclassical
large \(N\) expansion: the UV divergence produced by the near-horizon bulk
fields is of the same local form as the matter-loop renormalization of the
Einstein-Hilbert term, and can therefore be absorbed into the renormalized
Bekenstein--Hawking area term.  In this sense the continuum bulk EFT is a
consistent large-\(N\) description after renormalization.

The finite $N$ question is different.  Renormalization does not imply that
the continuum near-horizon Hilbert space of the bulk EFT is an exact
microscopic Hilbert space of the CFT.  It only says that the local UV
divergence is handled consistently order by order in the bulk EFT.  At
finite \(N\), the microscopic theory has a finite entropy \(S_{\rm BH}\) and,
on a compact boundary, a discrete spectrum.  Hence the large \(N\)
continuum approximation must eventually fail in sufficiently fine-grained
or late-time observables.  In the present paper the brick wall is used as a
phenomenological way to model this finite $N$ deviation from the
renormalized large \(N\) bulk expansion.

\paragraph{Localization of the energy and entropy}
The estimates above also show where the energy and entropy live.  Both integrals are dominated by the lower endpoint $\rho\sim \epsilon$.  More explicitly,
\begin{equation}
  dS\sim A_h\frac{d\rho}{\rho^{D-1}},
  \qquad
  dE
  \sim A_h\kappa\frac{d\rho}{\rho^{D-1}}.
\end{equation}
Thus the relevant degrees of freedom are localized in a Planckian proper layer around the stretched horizon.  Their local energy scale is Planckian, but the Killing energy seen at infinity is red-shifted.  This is why the same layer can simultaneously be a strongly interacting UV system locally and a thermal reservoir at the Hawking temperature for asymptotic observers.
This resolves a possible paradox.  A mode with $\omega\sim T_H$ appears low energy in the (asymptotic) Hamiltonian, but near $\rho\sim\ l_{p}$ it has local frequency of order the cutoff.  Interactions among such modes need not be weak.  Consequently, the Dirichlet wall is not the complete physical model.  The complete effective model consists of the brick-wall regulator plus the strongly interacting stretched-horizon degrees of freedom that the regulator is standing in for.

\paragraph{Thermal equilibrium and typical states}
For the Hartle--Hawking-like equilibrium problem, it is natural to describe
the stretched-horizon sector by a thermal density matrix at the Hawking
temperature, $\rho (T_H)$.
This is not merely a notational convenience: it is the state appropriate
to the equilibrium black hole, in which the exterior modes and the
stretched-horizon reservoir are in detailed balance at temperature \(T_H\).
One may also replace this thermal density matrix by a typical pure state
in the corresponding microcanonical energy window, provided one assumes
the usual statistical-mechanical typicality for the relevant coarse-grained
observables.  Namely, for a simple exterior "macro" observable \(O_{\rm simple}\)
which does not resolve exponentially fine microstate information,
\begin{align}
  \langle\Psi|O_{\rm simple}|\Psi\rangle
  \simeq
  {\rm Tr}(\rho_{\rm micro} O_{\rm simple})
  \simeq
  {\rm Tr}(\rho (T_H) O_{\rm simple}) .
\end{align}
This equivalence is not an additional microscopic derivation of the
Hartle--Hawking state; it is the usual statistical-mechanical statement
that typical pure states and thermal density matrices agree for ordinary
coarse-grained observables.


For the evaporating problem discussed below, the same reservoir is taken
to be thermally or typically populated, while the ordinary exterior
incoming modes are placed in their vacuum; this gives the brick-wall
analogue of the Unruh state.

\subsection{Effective dynamics: why an interacting brick wall can look black}
\label{s32}

A strictly free Dirichlet wall is perfectly reflecting.  If the field is
free, an incoming exterior one-particle wave returns as an outgoing
one-particle wave with a phase,
\begin{align}
  |\omega,\ell,s\rangle_{\rm ext}
  \longrightarrow
  e^{i\delta_{\ell s}(\omega)}
  |\omega,\ell,s\rangle_{\rm ext}.
\end{align}
This is the usual objection to interpreting a brick wall as a black hole
horizon.

The point changes once the near-wall region is treated as an interacting
system.  Near the stretched horizon, modes with small Killing energy are
locally cutoff-scale excitations.  Therefore gravitational and other
interactions in this layer need not be weak.  An incoming exterior wave can
scatter into many near-wall excitations, each carrying a small fraction of
the asymptotic energy.  The final state is then not a coherent outgoing
one-particle wave.  For an observer who measures only the ordinary exterior
one-particle sector, the probability has effectively disappeared into the
near-wall degrees of freedom.

This is the sense in which the interacting brick wall can look like a black
object, as in the mechanism discussed in \cite{Terashima:2025tct}.  The wall remains a
regulator of the near-horizon spectrum, but the inclusive response of the
exterior simple sector is absorptive.  The absorption is not fundamental
non-unitarity.  It is ordinary unitary evolution in the enlarged Hilbert
space, followed by coarse graining over the complicated near-wall final
states.

Thus the claim is simple.  A free brick wall reflects, but an interacting
brick wall can look approximately black to simple exterior probes.  The
coherent reflected wave can be strongly converted into many near-wall
degrees of freedom.  At finite \(N\), this conversion cannot be an exact
continuum absorption forever; the hidden unitary dynamics can appear as reflection, echoes,
or spectral fine structure.

\subsection{Hawking radiation as ordinary black-body emission from the stretched horizon}

The previous subsection explained why the near-wall reservoir can give an
absorptive response to exterior probes.  Absorption, however, does not by
itself determine the state of the reservoir.  To discuss Hawking radiation
one must also specify whether the exterior ordinary modes are thermally
populated or are in their vacuum.  This is the role of the
Hartle--Hawking-like and Unruh-like states in the brick-wall description.

A Hartle--Hawking-like state is the equilibrium state in which both the
near-wall reservoir and the ordinary exterior modes are thermal at
\(T_H\) with respect to the asymptotic time.  There is then no net
luminosity at infinity: outgoing and incoming thermal fluxes balance.

For the evaporating black hole, the relevant state is Unruh-like.  In the usual horizon QFT language one often describes this as a state in which the outgoing sector is thermally populated while the incoming sector is in its vacuum.  In a brick-wall model this left-moving/right-moving language is not intrinsic, because the boundary condition ties the two components into standing-wave normal modes.  
Instead, for the purpose of the effective coarse-grained description, we
approximately separate the degrees of freedom according to their physical
role:
\begin{align}
  \text{near-wall reservoir}
  \quad\oplus\quad
  \text{ordinary exterior modes}.
\end{align}
This should not be understood as an exact tensor-factor decomposition of
the microscopic Hilbert space.  The two sectors interact, and in gravity
there are also constraints and dressings which obstruct an exact local
factorization.  The statement is only that, for ordinary exterior probes
and for the time scales on which the stretched-horizon reservoir can be
treated as unobserved, this approximate separation is useful.
With this effective separation, the Unruh-like state is represented
schematically as
\begin{align}
  \rho_{\rm U-like}
  \simeq
  \rho_{\rm reservoir}(T_H)\otimes |0_{\rm ext}\rangle\langle 0_{\rm ext}|,
  \label{Ustate}
\end{align}
or, equivalently, with the thermal state, $\rho_{\rm reservoir}$, replaced by a typical pure state in the corresponding energy window.  The first factor means that the Planckian stretched-horizon layer is thermally or typically populated.  The second factor means that there is no incoming thermal bath in the ordinary exterior modes.

This is the brick-wall analogue of the Unruh vacuum.  It should not be described as assigning independent occupation numbers to left- and right-moving modes at the wall.  Rather, the stretched-horizon sector is hot, while the exterior is empty.  In an asymptotically flat spacetime this is precisely the physical setup of a hot object radiating into vacuum.

The interacting brick-wall model therefore gives Hawking radiation in the
same sense as an ordinary hot body gives black-body radiation: the reservoir
is thermally populated, the exterior is empty, and the emitted quanta are
filtered by the greybody potential.  This is not a derivation of Hawking
radiation from a free Dirichlet wall; it assumes that the interacting
near-wall reservoir has order-one emissivity for the relevant exterior
modes.\footnote{
Within the thermal description of the near-wall reservoir, its
equilibrium correlators obey the KMS condition and the usual
detailed-balance relation between absorption and emission.
Thus the effective absorption discussed above also gives thermal
emission at \(T_H\); an independent emission mechanism is not required.
The full Unruh-like state is not KMS, since the exterior incoming
modes are in their vacuum.
Propagation through the unchanged exterior potential gives the usual
greybody filtering of this thermal emission.
}


\paragraph{Summary of the effective picture}
The combined picture is therefore as follows.
\begin{enumerate}
\item
The brick-wall cutoff regulates the infinite continuum of near-horizon EFT
modes and replaces it by a large but finite stretched-horizon reservoir.

\item
The free WKB calculation explains why the entropy and Killing energy are
localized in a Planckian layer and scale as
\begin{align}
  S\sim {A_h\over l_{p}^{D-2}},
  \qquad
  E\sim T_H {A_h\over l_{p}^{D-2}} .
\end{align}
This establishes the parametric consistency with black hole
thermodynamics.

\item
A free Dirichlet wall is reflecting, but interactions with the dense
near-wall spectrum can convert the coherent exterior one-particle
reflection into many near-wall final states.  Thus the exterior simple
sector can see an absorptive, black hole-like response.

\item
For the Hartle--Hawking-like equilibrium problem, both the near-wall
reservoir and the ordinary exterior modes are thermal at \(T_H\).  For the
Unruh-like evaporating problem, the reservoir is thermally populated while
the ordinary exterior incoming modes are in their vacuum.  
In asymptotically flat spacetime this is just a hot object radiating into
vacuum.  If the interacting reservoir has order-one emissivity for the
relevant modes, the emitted quanta, after the usual greybody filtering,
give the Hawking flux.

\item
The model is thermodynamically consistent in the usual sense.  The
interacting reservoir may have an equation of state different from the
free brick-wall gas, but small variations obey \(dE=T_H dS\) on the
fixed background.  In the Unruh-like state, the emitted radiation carries
entropy outward, as for an ordinary hot body radiating into vacuum.
\end{enumerate}

In this sense, the brick-wall model with gravitational interactions is not
merely a state-counting device.  It is a consistent effective
stretched-horizon model of black hole thermodynamics, absorption, and
Hawking radiation, up to order-one coefficients whose values must be
determined by the microscopic theory.

\section{Retarded response, quasinormal modes, and echoes}\label{sec:response-echoes}

The brick-wall reflection is most cleanly stated in the retarded Green function.
In this section we explain this response-theory statement first; the finite $N$ interpretation of the same reflection is discussed in Section~\ref{sec:finiteN-reflection}.
We stress that the brick wall does not reproduce the black-hole pole spectrum, namely the quasinormal-mode spectrum.
One might then worry that this is in conflict with the standard gravitational-wave ringdown calculations based on quasinormal modes.
However, the brick-wall model reproduces the exterior black-hole response up to the first return time from the wall.
Thus there is no discrepancy for the early ringdown signal observed before any possible echo returns.


\subsection{Early-time quasinormal response from a reflecting brick wall}

In this subsection we clarify the sense in which the usual quasinormal-mode
description of the exterior response is reproduced in a brick-wall model.  The
statement is not that a reflecting brick wall has the same exact pole spectrum
as a black hole.  It does not.  The black hole problem uses an absorptive
ingoing boundary condition at the horizon, while a Dirichlet wall is perfectly
reflecting.  The appropriate statement is instead a statement about the
retarded Green function at finite time.

Consider a perturbation variable \(\Psi(t,r_*)\) obeying
\begin{equation}
 \left[
 -\partial_t^2+\partial_{r_*}^2-V_\ell(r)
 \right]\Psi(t,r_*)=0 ,
\end{equation}
where \(r_*\) is the tortoise coordinate and the horizon is at
\(r_*=-\infty\).  Let \(G_{\rm BH}^R\) denote the retarded Green function of
the black hole problem, defined by imposing the ingoing condition at the
horizon.  Let \(G_{\rm BW}^R\) denote the retarded Green function of the
brick-wall problem, in which the spacetime is truncated at
\begin{align}
 r_*=r_*^{\rm w}
\end{align}
and the Dirichlet condition
\begin{align}
 \Psi(t,r_*^{\rm w})=0
\end{align}
is imposed.  The outer boundary condition is taken to be the same in the two
problems.

Let the source and the observation point, $r_*',r_*$, lie in a fixed exterior region
\begin{align}
 r_*,r_*' > r_*^{\rm obs},
\end{align}
with \(r_*^{\rm obs}\) outside the near-horizon region.  Define the echo time
by
\begin{equation}
 t_{\rm echo}(r_*,r_*')
 =
 (r_*-r_*^{\rm w})+(r_*'-r_*^{\rm w}) ,
\end{equation}
or, for observations near the exterior potential barrier,
\begin{equation}
 t_{\rm echo}\simeq 2|r_*^{\rm w}-r_*^{\rm peak}| .
\end{equation}
Here \(r_*^{\rm peak}\) is the position of the main exterior potential barrier of the Black hole.
This is the earliest time at which a signal emitted from \(r_*'\), reflected at
the wall, can return to \(r_*\).

By finite propagation speed, the difference between the two retarded Green
functions vanishes before this reflected signal can arrive:
\begin{equation}
 G_{\rm BW}^R(t,r_*;t',r_*')
 =
 G_{\rm BH}^R(t,r_*;t',r_*')
 \qquad
 \text{for }
 0<t-t'<t_{\rm echo}(r_*,r_*') .
\end{equation}
This equality is the basic reason why the early-time exterior response is
insensitive to the presence of the wall.  The wall changes the retarded Green
function only after the first null ray has had enough time to travel to the
wall and back.

For a Planck-scale stretched horizon one has, parametrically,
\begin{equation}
 |r_*^{\rm w}|
 \sim
 r_h \log\frac{r_h}{\ell_{\rm P}} ,
\end{equation}
and therefore
\begin{equation}
 t_{\rm echo}
 \sim
 2r_h\log\frac{r_h}{\ell_{\rm P}} .
\end{equation}
For a macroscopic black hole this is much larger than the ordinary ringdown
time,
\begin{equation}
 t_{\rm ring}\sim O(r_h).
\end{equation}
Hence there is a parametrically large time window
\begin{equation}
 t-t' < t_{\rm echo}
\end{equation}
in which the brick-wall Green function is identical to the black hole Green
function in the exterior region.

In this window the black hole response may be approximated by the usual
quasinormal-mode expansion,
\begin{equation}
 G_{\rm BH}^R(t,r_*;t',r_*')
 \simeq
 \sum_j
 \mathcal R_j(r_*,r_*')
 e^{-i\omega_j(t-t')} ,
\end{equation}
where \(\omega_j\) are the black hole quasinormal frequencies.  Since
\(G_{\rm BW}^R=G_{\rm BH}^R\) before the echo time, the same quasinormal
expression also describes the brick-wall response in this time interval:
\begin{equation}
 G_{\rm BW}^R(t,r_*;t',r_*')
 \simeq
 \sum_j
 \mathcal R_j(r_*,r_*')
 e^{-i\omega_j(t-t')},
 \qquad
 r_h\lesssim t-t'<t_{\rm echo}.
\end{equation}
Thus the standard quasinormal-mode calculation, such as the one used to model
the early gravitational-wave ringdown, is reproduced by the brick-wall model
up to the echo time.

The difference appears only at later times.  It can be written schematically as
an echo expansion,
\begin{equation}
 G_{\rm BW}^R(t,r_*;t',r_*')
 =
 G_{\rm BH}^R(t,r_*;t',r_*')
 +
 \sum_{m=1}^{\infty}
 G_{\rm echo}^{(m)}(t,r_*;t',r_*') ,
\end{equation}
where \(G_{\rm echo}^{(m)}\) represents a contribution involving \(m\)
reflections from the wall and satisfies
\begin{equation}
 G_{\rm echo}^{(m)}(t,r_*;t',r_*')=0
 \qquad
 \text{for }
 t-t'<m\,t_{\rm echo}.
\end{equation}
The first correction is therefore delayed until \(t-t'\simeq t_{\rm echo}\).
This delayed contribution is the echo signal.


The same statement has a simple frequency-space interpretation.  A
contribution delayed by a time \(t_{\rm echo}\) gives, after Fourier
transformation, an oscillating factor \(e^{i\omega t_{\rm echo}}\).  Thus
the difference between the brick-wall and black hole Green functions has
the schematic form
\begin{align}
  \Delta G^R(\omega,\ell)
  \equiv
  G^R_{\rm BW}(\omega,\ell)-G^R_{\rm BH}(\omega,\ell)
  \sim
  e^{i\omega t_{\rm echo}}F_\ell(\omega)
  +\text{higher echoes},
\end{align}
where \(F_\ell(\omega)\) contains the ordinary greybody and
source/observer dependence and is smooth on the fine frequency scale
\(t_{\rm echo}^{-1}\).  Hence the brick-wall answer contains oscillatory
fine structure with spacing
\begin{align}
  \Delta\omega_{\rm BW}\sim {1\over t_{\rm echo}},
\end{align}
up to order-one factors.  If the Green function is probed with frequency
resolution
\begin{align}
  \delta\omega \gg {1\over t_{\rm echo}},
\end{align}
this fine structure is averaged over, and one recovers the coarse-grained
black hole response,
\begin{align}
  G^R_{\rm BW}(\omega,\ell)\simeq G^R_{\rm BH}(\omega,\ell).
\end{align}

In summary, a Dirichlet brick wall does not reproduce the black hole
quasinormal spectrum as a set of complex poles.  Nevertheless, it reproduces
the exterior black hole retarded Green function for all times shorter than the
echo time.  Consequently, the usual quasinormal-mode description of the
early-time ringdown is reproduced by the brick-wall model.  The deviation from
the black hole answer appears only at later times as echoes, or equivalently
in frequency space as a rapidly oscillating fine structure with spacing
\(\Delta\omega_{\rm BW}\sim 1/t_{\rm echo}\).

\subsection{Echoes and damping from the same reservoir}

The damping of echoes is the time-domain counterpart of the coarse-grained
absorption discussed in subsection~\ref{s32}.  A reflecting structure outside the
horizon produces echoes: an outgoing wave is partially reflected by the
exterior angular-momentum barrier, returns to the near-wall region, and is
reflected again.  The characteristic delay is twice the tortoise-coordinate
separation between the near-wall region and the exterior potential barrier,
\begin{align}
  t_{\rm echo}\simeq 2|r_*(r_{\rm wall})-r_*(r_{\rm barrier})| .
\end{align}
For a Schwarzschild-like geometry this gives the familiar logarithmic estimate
\begin{equation}
  t_\mathrm{echo}
  \sim \kappa^{-1}\ln\frac{r_h}{\epsilon},
\end{equation}
up to order-one constants and convention-dependent details.  If the wall is strictly reflecting,
these echoes are long-lived, limited only by leakage through the exterior barrier.

In the reservoir picture, each encounter with the near-wall region has an inclusive probability
$\mathcal A_{\ell s}(\omega)$ of losing the coherent one-particle amplitude into many near-wall
excitations carrying small fractions of the asymptotic energy.  Therefore, in a simple
phenomenological parametrization, the echo amplitude is multiplied by approximately
\begin{equation}
  \sqrt{1-\mathcal A_{\ell s}(\omega)}
\end{equation}
per encounter, in addition to the usual greybody factors.  This gives a physical interpretation of
stretched-horizon damping: the echo is not necessarily absorbed by a fundamental hard surface; it is not simply reflected, but dephased and distorted, because the incoming coherent wave is scrambled into the near-wall degrees of freedom.\footnote{
The echo time is the same as the scrambling time \cite{Sekino:2008he}, with an ambiguous ${\cal O}(1)$ overall factor.
Here, we think the coherent wave is scrambled and reflected at the same time,
as argued in \cite{Terashima:2025tct}.
}

This section should not be read as a precise prediction for gravitational-wave echoes.  The echo
amplitude, phase, and delay depend on the effective near-wall response, on the exterior potential
barrier, and on the state of the reservoir.  The point is instead qualitative: 
the same coupling of exterior modes to many near-wall excitations that gives
effective dissipation also damps echoes.
Thus the reservoir picture connects
three notions which otherwise appear separate: brick-wall state counting, stretched-horizon
absorption, and the damping of the echo.

\begin{table}[ht]
\centering
\renewcommand{\arraystretch}{1.4}
\begin{tabular}{@{}p{0.30\textwidth} p{0.20\textwidth} p{0.44\textwidth}@{}}
\hline
\textbf{Probe / channel} & \textbf{Time scale} & 
\textbf{Effective exterior response} \\
\hline
Generic high-energy probe ($\omega \gg T_H$), 
e.g.\ ordinary matter and light 
& within the coarse-grained regime 
& Coherent reflection strongly depleted into the near-wall reservoir, 
suppressed by $\sim\exp(-c\,\omega^2/T_H^2)$. The exterior simple sector 
sees an absorptive, black hole--like response. \\
\hline
Low-energy / gravitational channel ($\omega \sim T_H$), 
& $t - t' < t_{\rm echo}$ 
& $G^R_{BW} = G^R_{BH}$ exactly, by causality: the reflected signal has not 
yet returned. The standard quasinormal ringdown is reproduced. \\
\hline
Low-energy / gravitational channel ($\omega \sim T_H$), 
& $t - t' \gtrsim t_{\rm echo}$ 
& Residual return (echo). Not $e^{-S_{\rm BH}}$--suppressed but of order $N^0$, 
partially reduced by the same $\exp(-c\,\omega^2/T_H^2)$ factor and by 
greybody filtering; appears on the logarithmic scale 
$t_{\rm echo}\sim\kappa^{-1}\log S_{\rm BH}$. \\
\hline
\end{tabular}
\caption{Effective exterior response of the interacting brick wall, organized 
by probe energy and by time scale. The semiclassical ``perfect absorber'' 
picture is recovered for ordinary probes within the coarse-grained regime, and for the 
gravitational channel before the echo time; the only departure is the echo, 
which is an $O(N^0)$ effect on the logarithmic stretched-horizon time scale 
rather than an $e^{-S_{\rm BH}}$ effect on the Heisenberg time scale.}
\label{tab:channels}
\end{table}

In the next section, we will argue that 
there can be an $O(N^0)$ return by the finite $N$ effects.
Table~\ref{tab:channels} shows why the success of the semiclassical 
description and the survival of an $O(N^0)$ return are not in tension. They 
live in different channels and on different time scales: ordinary probes see 
a black object, the gravitational ringdown agrees with the smooth-horizon 
answer up to the echo time, and the finite $N$ departure is confined to the 
echo.

\section{Finite $N$ thermal correlators and unavoidable reflection}
\label{sec:finiteN-reflection}

This section states the main finite $N$ lesson of the brick-wall
description.  A brick wall placed outside the horizon necessarily reflects:
this is obvious at the level of the regulated bulk wave equation, because the
near-horizon continuum has been replaced by a finite interval with a boundary
condition.  The claim of this section is that this reflection should not be
regarded as an artifact of the brick-wall regulator.  Rather, it is the
brick-wall manifestation of a general property of finite $N$ AdS/CFT,
viewed as a complete theory of quantum gravity.  A finite $N$ black hole in
AdS is described by an ordinary quantum system with a discrete spectrum on a
compact boundary space.  Such a system cannot be exactly equivalent, for all
times and all correlators,\footnote{
The logarithmic time scale discussed below should not be confused with the
Heisenberg time of the exact finite $N$ spectrum. The latter is the time
needed to resolve individual many-body levels and is exponentially long in
the entropy. The time scale relevant for the present discussion has a
different origin. 
We consider low-energy, simple exterior perturbations, for which
semiclassical bulk EFT remains reliable in the exterior region outside a
Planckian stretched horizon.
Then finite $N$ effects cannot substantially modify propagation in this exterior EFT
region. At the same time, finite \(N\) cannot support the ideal continuum of
near-horizon EFT modes all the way to the mathematical horizon. The failure
of the continuum description must therefore be localized in the
stretched-horizon region.
A perturbation sent from the exterior first becomes sensitive to this
finite $N$ region only after it has propagated to the stretched horizon
and back. Because of the near-horizon redshift, this causal time is
$
t_{\rm sh}\sim \kappa^{-1}\log {r_h\over l_{p}}
\sim \kappa^{-1}\log S_{\rm BH}.
$
In the brick-wall effective model the corresponding finite $N$ correction
appears as a reflection or echo. In the exact finite $N$ CFT it need not
be a clean geometric echo, but the same reasoning suggests that the first
correction to the ideal continuum-absorber approximation should occur on
this logarithmic stretched-horizon time scale, rather than only at the
Heisenberg time.
}
to a description in which the ordinary thermal or one-sided Hamiltonian
evolution is identified with the smooth Kruskal evolution and the
horizon acts as an irreversible sink in the reduced description.

Thus, the brick wall should be understood as a simple, effective way of making
this finite $N$ obstruction visible.  In the semiclassical
bulk EFT based on \(N=\infty\), an incoming perturbation can be described as passing through the
horizon, and exterior retarded response functions are computed with purely
ingoing boundary conditions.  In a finite $N$ theory, however, there is no
fundamental continuum sink in the exact microscopic dynamics.  The signal may
be delayed, scrambled, and strongly suppressed, and it need not resemble a
prompt reflection from a hard mirror.  Nevertheless, some residual return of
information is unavoidable.  In the brick-wall model, this return appears as
reflection or echoes; in the exact CFT language, it appears as 
late-time correlations and departures from the correlator
obtained by making an identification between the ordinary thermal
evolution and the smooth Kruskal evolution.\footnote{
For Rindler physics with a UV Cutoff on a lattice, which can be regarded as a toy model of the finite $N$ bulk model of a horizon, the claims in this section were observed in \cite{Chikazawa:2026iro}.
}

We now state the logic of this section, since it determines the status
of the return.  The stretched horizon is not an assumption that we add
to the finite \(N\) CFT; its existence, and the breakdown of the
equivalence principle there, are forced by combining the following
facts.
First, finite \(N\) cannot support the near-horizon continuum of the
semiclassical EFT: the entropy of the exterior EFT modes already
saturates \(S_{\rm BH}\) at a Planckian proper distance from the
horizon (Sections~\ref{sec:finiteN-viewpoint}
and~\ref{sec:state-counting}, and \cite{Iizuka:2013kma}).  The
continuum description must therefore end there.
Second, the smooth Kruskal evolution through the horizon is not the
evolution generated by the decoupled Hamiltonians \(H_L\pm H_R\).
Such a horizon-crossing evolution cannot in general be represented,
for arbitrary two-sided initial states, by a state-independent
Hamiltonian acting on the fixed right-CFT algebra
(subsection~\ref{subsec:nogo}).  The prediction of the semiclassical
EFT that a packet crosses the horizon and disappears is therefore not a
prediction of the finite \(N\) CFT with its own Hamiltonian; the two
evolutions differ.
Third, the exterior bulk EFT, with its order-one interactions, is valid
in the exterior region down to a Planckian proper distance.  The
difference between the finite \(N\) evolution and the smooth-horizon EFT
cannot be located in the exterior, where the EFT is reliable; it must be
located where the EFT ends, at the stretched horizon.
Taken together, these facts imply that the equivalence principle fails
at the stretched horizon in the finite \(N\) CFT: the exterior EFT modes
terminate on a surface a Planckian distance outside the horizon, and
what happens there is not free propagation into an interior.  This is a
consequence of finite \(N\) and of the no-go, not a model assumption.
To estimate the exterior return, we use the effective near-horizon
dynamics discussed below. A packet reaching the stretched horizon can
be converted into stretched-horizon excitations by the order-one
interactions there. Causality fixes the earliest time at which the
exterior can respond to this dynamics, namely the stretched-horizon
time \(t_{\rm echo}\sim\kappa^{-1}\log S_{\rm BH}\).
We expect a residual return on this logarithmic time scale, with an
amplitude controlled by the near-horizon interactions and the energy
of the incident packet, as estimated in
subsection~\ref{subsec:unavoidable-reflection}, rather than by the
many-body level spacing. The expected \(O(N^0)\) size is a dynamical
estimate; it is not fixed by causality or by the order of the
interaction strength alone.
The brick-wall model is the simplest effective model of this necessary
breakdown of the equivalence principle at the stretched horizon.  It is
used below to make the return explicit and to estimate its size. The
argument above motivates the return, while its logarithmic time scale
and expected \(O(N^0)\) amplitude are estimated using the effective
near-horizon dynamics.

\subsection{No-go for a Kruskal-like time evolution for a single CFT}
\label{subsec:nogo}

We will consider the (different) time evolutions for the eternal black hole in AdS/CFT.  
The ordinary
TFD state is a state in
\begin{align}
        {\cal H}_L\otimes {\cal H}_R
\end{align}
of two decoupled CFTs on $S^{d-1}$ where $d=D-1$.  
The standard boundary time translations are generated by
\(H_L+H_R\), or, depending on convention, by the boost generator \(H_R-H_L\),
where $H_L, H_R$ are represented by the Dilatation operators.
These Hamiltonians generate the usual asymptotic boundary time translations.
In the usual semiclassical interpretation, however, this ordinary TFD
evolution is also taken to describe the eternal black hole geometry with a
smooth interior \cite{Maldacena:2001kr}.  Different bulk slicings of that geometry, including
Kruskal-like slicings through the horizon, are then regarded as
diffeomorphic descriptions of the same bulk physics within the EFT.  In this
sense, the standard intuition implicitly identifies the ordinary TFD
Hamiltonian evolution with the smooth Kruskal evolution, at least for the
observables and time scales described by the semiclassical bulk EFT.

The point emphasized here is that this identification is an assumption of the
bulk EFT description; it is not a consequence of the finite $N$ CFT.
Diffeomorphism invariance is a property of the emergent bulk EFT.  If the EFT
is pushed beyond its regime of validity, its diffeomorphism-based
identification of different bulk time evolutions need not remain an exact
microscopic statement.  To make this distinction explicit, let \(H_{\rm Kr}\)
denote a Hamiltonian on \({\cal H}_L\otimes {\cal H}_R\) that realizes the
smooth Kruskal-like evolution.  There is no reason, in the exact finite $N$
CFT, to require
\begin{align}
        H_{\rm Kr}=H_L+H_R,
        \qquad
        H_{\rm Kr}=H_R-H_L .
\end{align}
More generally, if the Kruskal evolution is realized as a Hamiltonian
evolution on the two-CFT Hilbert space, it can be a genuinely different
Hamiltonian from the decoupled TFD generator.  Schematically,
\begin{align}
        H_{\rm Kr}=H_L+H_R+H_{\rm mix}^{\rm Kr},
\end{align}
where \(H_{\rm mix}^{\rm Kr}\) denotes the non-factorized part in the
left--right tensor product.\footnote{
There is a simple alternative possibility.  It may be that no microscopic
Hamiltonian \(H_{\rm Kr}\) exists which implements the smooth Kruskal-like
evolution on the finite \(N\) CFT Hilbert space.  This possibility is not an
objection to the present argument.  On the contrary, it would mean that the
Kruskal time evolution is only a notion inside the semiclassical bulk EFT
and has no exact counterpart in the finite \(N\) quantum theory.
Thus there are two possibilities.  If a microscopic \(H_{\rm Kr}\) exists, it
need not be equal to \(H_L+H_R\) or \(H_R-H_L\), and in general it would have
to act nontrivially on the two sides.  If no such \(H_{\rm Kr}\) exists, then
the smooth Kruskal evolution itself is absent as a microscopic time
evolution.  In either case, the usual semiclassical identification of
Kruskal evolution with the ordinary TFD Hamiltonian evolution is not a finite \(N\) statement.
} 
We will show this below.

As discussed in \cite{Leutheusser:2021frk}, the decoupled generators
\(H_\pm=H_R\pm H_L\) leave a right-CFT operator in the right-CFT algebra:
\begin{align}
{\cal O}_R(t)
&=e^{iH_\pm t}{\cal O}_R e^{-iH_\pm t}
 =e^{iH_R t}{\cal O}_R e^{-iH_R t},\\
[{\cal O}_R(t),{\cal O}_L(t')]&=0
\qquad\text{for all }t,t'.
\end{align}
Here \({\cal O}_L(t')=e^{iH_Lt'}{\cal O}_L e^{-iH_Lt'}\) belongs to the
left-CFT algebra. This does not imply the absence of left--right
correlations in the TFD state. The distinction concerns unequal-time
commutators. A Kruskal-like horizon-crossing flow can give
\[
[U_{\rm Kr}^{\dagger}(s){\cal O}_R(0)U_{\rm Kr}(s),
 {\cal O}_L(0)]\neq0,
\qquad U_{\rm Kr}(s)=e^{-iH_{\rm Kr}s},
\]
for suitable operators when the corresponding bulk observables become
causally related. This is exactly what is found in
\cite{Leutheusser:2021frk} for the emergent large-\(N\) infalling
evolution \(U(s)\): the commutator of the evolved right-exterior field
with a fixed left-exterior field becomes nonzero precisely when the
support of the evolved operator enters the light cone of the left
insertion, which is possible because \(U(s)\) involves the degrees of
freedom of both CFTs.
The left operator here represents an operation on
the initial slice, not a local operator on the later slice of the right
measurement. Equal-time commutators remain zero even in the presence of
\(H_{\rm mix}^{\rm Kr}\): evolving both operators by the same unitary
preserves their initial commutator. It is the unequal-time causal
response that the decoupled evolution cannot implement.
Now ask whether a Kruskal-like evolution can be represented purely within
the right CFT after the left CFT has been traced out.  Given an initial
two-sided state \(\rho_{LR}\), the right-reduced state obtained after a
Kruskal-like evolution is
\begin{align}
  \rho_R^{\rm Kr}(t)
  =
  {\rm Tr}_L\!\left[
    e^{-iH_{\rm Kr}t}\rho_{LR}e^{iH_{\rm Kr}t}
  \right].
\end{align}
In general, this is not equal to the right-CFT evolution
\begin{align}
  \rho_R^{\rm Kr}(t)
  \neq
  e^{-iH_Rt}\rho_R(0)e^{iH_Rt},
  \qquad
  \rho_R(0)={\rm Tr}_L\rho_{LR}.
\end{align}
If \(H_{\rm Kr}\) contains the non-factorized part \(H^{\rm Kr}_{\rm mix}\),
then the reduced right-side dynamics is generally an open-system evolution,
not a unitary evolution generated by \(H_R\) or by any Hermitian
Hamiltonian on \({\cal H}_R\).

More strongly, the exact reduced time evolution is not
captured by simply replacing the Hamiltonian by a non-Hermitian one.  Tracing
out an interacting environment gives reduced dynamics with noise,
decoherence, and, in general, memory effects. For a genuinely
non-factorizing evolution, the reduced dynamics cannot in general be
reproduced, for arbitrary initial two-sided states, by a single
state-independent Hermitian Hamiltonian acting on the fixed right-CFT
algebra.

The standard TFD Hamiltonians \(H_L\pm H_R\) generate the
usual boundary time translations of the two decoupled CFTs.  In the
semiclassical bulk EFT they are normally interpreted as giving the same
eternal black hole physics as a smooth Kruskal description.  The claim here is
that this equality is not valid at finite \(N\).  If one first evolves with
\(H_{\rm Kr}\) and then traces out the left CFT, the resulting right-side
dynamics is generically an open-system evolution, not the unitary evolution
generated by \(H_R\) or by any Hermitian Hamiltonian on
\({\cal H}_R\).

Thus, the smooth Kruskal-like evolution cannot be identified
with the evolution generated by \(H_R\) on the fixed
right-CFT algebra.\footnote{
The TFD state evolved by \(H_L+H_R\) and the same TFD state evolved by \(H_{\rm Kr}\), if the latter exists, define different dynamical systems even though they share the same initial state.
}
This is a failure of the microscopic equivalence suggested by
the semiclassical bulk description.\footnote{
The works
\cite{Leutheusser:2021frk,Leutheusser:2021qhd,Leutheusser:2022bgi}
construct emergent infalling evolutions in the large-\(N\) framework.
The relevant thermal single-trace algebras have an emergent
type III\(_1\) structure, and their GNS Hilbert space does not
admit the left--right tensor-product factorization of the
microscopic two-CFT theory. The resulting flow can generate
the unequal-time commutators discussed above.
Their explicit construction of the infalling evolution is carried
out at leading order in \(1/N\), in the non-factorizing GNS
representation described above. The identification of this
emergent evolution with ordinary CFT Hamiltonian evolution in
the exact finite-\(N\) theory, whose microscopic two-CFT Hilbert
space is \(\mathcal H_L\otimes\mathcal H_R\), is not analyzed there.
Our claim concerns this microscopic identification on the fixed
CFT algebras, even if a Kruskal-like implementation exists.
}

\begin{figure}[t]
\begin{center}
\begin{tikzpicture}[
  font=\footnotesize,
  >=Latex,
  boundary/.style={very thick},
  horizon/.style={dashed, thick},
  singularity/.style={very thick, decorate,
    decoration={snake, amplitude=0.7mm, segment length=3.0mm}},
  kruskal/.style={very thick},
  schw/.style={very thick, gray},
  lab/.style={align=center}
]


  \fill[gray!5]  (-3,-3) rectangle (3,3);
  \fill[gray!12] (-3,3) -- (0,0) -- (3,3) -- cycle;   
  \fill[gray!12] (-3,-3) -- (0,0) -- (3,-3) -- cycle; 
  \fill[gray!3]  (-3,-3) -- (-3,3) -- (0,0) -- cycle; 
  \fill[gray!3]  (3,-3) -- (3,3) -- (0,0) -- cycle;   

  \draw[boundary] (-3,-3) -- (-3,3);
  \draw[boundary] ( 3,-3) -- ( 3,3);

  \draw[singularity] (-3,3) -- (3,3);
  \draw[singularity] (-3,-3) -- (3,-3);

  \draw[horizon] (-3,-3) -- (3,3);
  \draw[horizon] (-3,3) -- (3,-3);

  \fill (0,0) circle (1.4pt);

  \node[lab, rotate=90] at (-3.38,0) {left AdS\\boundary};
  \node[lab, rotate=-90] at (3.38,0) {right AdS\\boundary};

  \node[lab] at (-2.05,0) {left\\exterior};
  \node[lab] at (2.05,0) {right\\exterior};

  \node[lab] at (0,1.85) {black hole\\interior};
  \node[lab] at (0,-1.75) {white-hole\\interior};

  \node[lab] at (0,3.38) {future singularity};
  \node[lab] at (0,-3.38) {past singularity};

  \node[lab, fill=white, inner sep=1pt] at (0,-0.65) {bifurcation\\surface};



  \draw[schw]
    (-3,1.05)
      .. controls (-2.55,0.98) and (-1.65,0.63) .. (-0.15,0.05);

  \draw[schw]
    (3,1.05)
      .. controls (2.55,0.98) and (1.65,0.63) .. (0.15,0.05);

  \node[lab, gray] at (0,-4.05)
    {gray curves: exterior Schwarzschild-time slices generated separately by $H_L$ and $H_R$};


  \draw[kruskal]
    (-3,1.05)
      .. controls (-2.35,0.48) and (-1.45,0.55) .. (-0.55,0.58)
      .. controls (-0.18,0.60) and (0.18,0.60) .. (0.55,0.58)
      .. controls (1.45,0.55) and (2.35,0.48) .. (3,1.05);

  \node[lab] at (0,4.05)
    {black curve: one smooth AdS--Kruskal-like Cauchy slice at $t_{\rm Kr}>0$};

  \draw[->, thick] (-3.85,-2.3) -- (-3.85,2.3)
    node[midway,left] {$t_L$};

  \draw[->, thick] (3.85,-2.3) -- (3.85,2.3)
    node[midway,right] {$t_R$};

\end{tikzpicture}
\end{center}
\caption{Two-sided eternal AdS black hole.  The conformal diagram is a square:
the vertical sides are the two timelike AdS boundaries, the horizontal sides
are the future and past singularities, and the two diagonals are the horizons.
The gray curves indicate Schwarzschild-like exterior time slices generated
separately by the two boundary Hamiltonians.  The black curve represents a
single Kruskal-like Cauchy slice,
connecting the two AdS boundaries through the Einstein-Rosen bridge.  
}
\label{fig:eternal-adsbh-kruskal-slicing}
\end{figure}

Fig. \ref{fig:eternal-adsbh-kruskal-slicing} shows an example of the two slices in the eternal black hole. 
A smooth Kruskal Cauchy slice is
not the union of two independently evolved exterior Schwarzschild slices.  It
passes through the interior as a single smooth hypersurface.  Within the
semiclassical bulk EFT, this difference is normally treated as a difference of
description, because gravity is diffeomorphism invariant.  Once translated to
the exact finite $N$ CFT, however, time evolution is specified by a
Hamiltonian.  Therefore the statement that the ordinary TFD Hamiltonian
evolution is the same as the smooth Kruskal evolution is an additional
identification, not a microscopic identity forced by the CFT definition.
The place where the two descriptions differ may be the bifurcation
surface.  
In the EFT for $H_L +H_R$, an infinite tower of modes accumulates
there; this is the same near-horizon continuum whose entropy diverges in
the brick-wall calculation of Section~\ref{sec:state-counting}.  Finite
\(N\) cannot support this continuum, so it is at the
bifurcation surface that the EFT ceases to be an exact
description, and the identification of its evolution with the
decoupled TFD evolution ceases to hold.

\subsection{Residual reflection at finite \(N\)}
\label{subsec:unavoidable-reflection}

Here, we consider a black hole described by the thermal state of the
right CFT evolved by \(H_R\), and argue that a wave packet sent toward the
would-be horizon experiences residual reflection, which is absent from
the smooth-horizon Kruskal-like EFT description.  (This can be regarded
as an analysis of thermal correlators.)

We first make clear where the two descriptions differ.  At \(N=\infty\),
the \(H_R\) evolution of the TFD state reproduces the smooth-horizon
correlators for right-exterior operators: this is the standard
thermofield construction \cite{Maldacena:2001kr}, and in this limit an
infalling packet simply disappears without returning.  The
difference between the \(H_R\) evolution and the Kruskal-like evolution
is then confined to left--right correlators, through the horizon-crossing
flow of the previous subsection; it is invisible to right-exterior
observables.
At finite \(N\), the right CFT cannot support the infinite tower of
near-horizon modes, and the exterior modes terminate at a stretched
horizon; as argued at the beginning of this section, this is forced by
finite \(N\) and the no-go, and the brick wall is only its simplest
model.  
In the Kruskal EFT, the
degrees of freedom on the two sides of the bifurcation surface are
connected by an order-one coupling, since a field propagates freely
across the surface.  If the EFT is regularized on a lattice with
Planckian spacing, as in \cite{Chikazawa:2026iro}, this coupling is
simply the ordinary kinetic term linking neighboring sites across the
surface.
In the finite \(N\) right-CFT description
this coupling is absent, because there is no left system for the
exterior modes to propagate into.  This is the content of
\(H^{\rm Kr}_{\rm mix}\) in the near-horizon region: an order-one local
coupling present in the Kruskal EFT and absent in the
\(H_R\) evolution.

A missing order-one coupling at the stretched horizon changes the
local generator there at order one. To infer the exterior response,
we also use the effective near-horizon dynamics and the wave-packet
estimate below, including conversion into stretched-horizon
excitations as discussed in subsection~\ref{s32}.
Information retained in the right CFT need not remain in a simple
exterior wave packet: it can be redistributed among complicated states
of the same stretched-horizon degrees of freedom. Information
retention alone therefore fixes neither the exterior return time nor
its amplitude. Causality gives the earliest possible response time
\(t_{\rm echo}\). Our expectation of a residual return on this time
scale, not suppressed as \(e^{-S_{\rm BH}}\), is instead based on the
effective local dynamics and the interaction estimate below.

The brick-wall model gives a particularly simple effective representation of
this finite $N$ obstruction.  In the free-field problem a wall placed a
proper distance \(\epsilon\) outside the horizon converts the continuum of
near-horizon modes into a discrete set of normal modes.  For times shorter
than the echo time, exterior correlators can agree with the black hole
retarded correlator, because the causal signal reflected by the wall has not
yet returned to the exterior observation region.  At later times, however,
the wall produces an echo.  In the phenomenological interpretation advocated
in this paper, the wall should not be viewed as a literal Planckian mirror.
Rather, it is a simple effective way of representing the fact that the
finite $N$ near-horizon sector is not an exact continuum.  The echo is
therefore the brick-wall realization of a more general finite $N$ statement:
perfect absorption is an idealization, while exact finite $N$ dynamics must
retain some memory of the excitation.

The size and detailed form of the returning signal are model-dependent.
Interactions in the near-horizon region can scramble the excitation, spread it
over many microscopic degrees of freedom, and make the return highly
incoherent.  Therefore, the general finite-$N$ conclusion is that thermal correlators may not always exhibit a clean, sharply resolved geometric echo of order one.

\paragraph{Estimate of the return amplitude}
We can ask about the expected size of this finite $N$ return.  
For an incoming wave packet, the effect of gravitational and other
interactions before the packet reaches the stretched horizon was estimated
in \cite{Terashima:2025tct}.  The conclusion is that the suppression
of the coherent reflected component is an \(O(N^0)\) effect, 
which may be understood, in a very simplified manner, as being due to the fact that the gravitational interaction becomes of order unity near the stretched horizon.
More
specifically, except for very low-energy packets, the coherent reflection
can be strongly suppressed, for example by a factor of the form
$
  \exp\!\left(-c\,{\omega^2\over T_H^2}\right),
$
with \(c=O(1)\).  
This suppression does not depend on the Planck scale and is independent of $N$ at fixed \(\omega/T_H\);
it is controlled by the energy of the wave packet in Hawking-temperature
units.

In the ideal free brick-wall problem the wall itself is perfectly
reflecting.  Thus, after the interaction-induced suppression of the
coherent one-particle component, the remaining reflected signal is of
order \(N^0\) in the large-\(N\) expansion, not nonperturbatively small
as \(e^{-S_{\rm BH}}\). We expect the same qualitative behavior for
finite \(N\) holographic CFTs.
The brick-wall counting of Section~\ref{sec:state-counting} accounts
parametrically for the \(O(N^2)\sim S_{\rm BH}\) degrees of freedom of
the black hole. This agreement in state counting does not by itself
fix the return amplitude: an excitation can be redistributed among
complicated many-body states of these same degrees of freedom.
Our expectation instead uses the effective local near-horizon
dynamics together with the wave-packet interaction estimate above.
Within this description, even without a sharp Dirichlet cutoff, we
expect an \(O(N^0)\) return at fixed \(\omega/T_H\).
The suppression in this estimate is controlled by the near-horizon
interaction and by the energy of the incident wave packet in
Hawking-temperature units, rather than by the many-body level spacing.
This is the basis for expecting a non-exponentially suppressed return
on the logarithmic stretched-horizon time scale, rather than for
inferring its magnitude from finite entropy alone.

This estimate should be regarded as qualitative.  It uses bulk EFT and the
stretched-horizon description up to a region where Planckian physics and
strong interactions are important.  Therefore the precise amplitude, phase,
and waveform of the returning signal cannot be determined without the
microscopic finite $N$ dynamics.  The important point is that the
finite $N$ reflection is expected to be an \(O(N^0)\) effect.  If such a return
survives in realistic gravitational perturbations, it could in principle
be probed by future gravitational-wave observations, as discussed in
\cite{Cardoso:2016rao, Abedi:2016hgu, Abedi:2020ujo} and
\cite{Terashima:2025tct} in the context of this paper.

In the reflecting brick-wall description, an \(O(N^0)\) echo is
compatible with the \(1/N\) expansion because \(t_{\rm echo}\) diverges
as \(N\to\infty\): at any fixed time, the causal support condition
makes the echo absent to all orders in \(1/N\).
This illustrates the non-uniformity of the large-\(N\) and late-time
limits. A related but distinct phenomenon occurs in the scrambling
analysis of \cite{Shenker:2013pqa}, where a correction to normalized
two-sided correlators grows from \(O(1/N^2)\) at fixed time to
\(O(1)\) on a logarithmic time scale. Unlike the pre-echo response
in the reflecting-wall problem, that correction is already present
at finite order in \(1/N\) at fixed time. A direct relation between
that effect and the one-sided return discussed here remains to be
established.

We also note that the relevant exterior spectrum can be continuous even at finite \(N\).  In the AdS--Rindler setup, the
Schwarzschild-like description uses CFTs on \(H^{d-1}\), and \(H^{d-1}\) is
non-compact.  Similarly, planar AdS black holes correspond to thermal states
on \(\mathbb{R}^{d-1}\).  In such cases the continuum spectrum can make the
thermal description effectively absorbing without requiring a finite-volume
recurrence.  This is why the usual thermal two-point functions on
\(\mathbb{R}^{d-1}\), and in particular on \(\mathbb{R}\) for \(d=2\), can
decay in the standard way.

\section{Near-extremal throats and finite $N$ state counting}
\label{sec:near-extremal}

The discussion so far was phrased in a Schwarzschild-like near-horizon
region, where the semiclassical geometry is well approximated by a Rindler
wedge outside the stretched horizon.  For an ordinary non-extremal black hole,
this is sufficient for the usual brick-wall state counting.  Near-extremal
charged black holes introduce an additional issue.  Their near-horizon region
contains a long \(AdS_2\)-like throat, whose large optical length gives a
dense spectrum of low-energy bulk EFT modes.

The point of this section is to determine when the thermal excitations of
these modes can be treated on a fixed classical near-extremal background.
We first recall the WKB density of states, then compare their thermal
Killing energy with the black hole's classical energy above extremality.
This gives a thermodynamic consistency scale for the throat description,
rather than a direct bound on one-particle mode number from the microscopic
Hilbert-space dimension.

We use units in which the asymptotic AdS scale and the horizon radius are of
order one,
\begin{equation}
        L_{\rm AdS}\sim r_h\sim 1 .
\end{equation}
The black hole entropy is denoted by
\begin{equation}
        S_{\rm BH}\sim {1\over G_N}\sim N_{\rm eff}^2 .
\end{equation}
For ordinary adjoint large-\(N\) theories, \(N_{\rm eff}^2\sim N^2\), while in
other holographic examples \(N_{\rm eff}^2\) should be understood as the
appropriate entropy scale.

\paragraph{The near-\(AdS_2\) throat and its optical length}

A near-extremal charged black hole has a near-horizon region of the form
\begin{align}
        ds^2 \simeq L_2^2\left[
        -(\rho^2-\rho_0^2)dt^2
        + {d\rho^2\over \rho^2-\rho_0^2}
        \right]
        + r_h^2 d\Sigma_{d-1}^2 .
\end{align}
Here \(\rho\) is a dimensionless radial coordinate in the near-\(AdS_2\)
throat.  The mouth of the throat is at \(\rho\sim O(1)\).  The extremal
geometry corresponds to \(\rho_0=0\), while the near-extremal outer horizon
is at
\begin{equation}
        \rho=\rho_0 ,
        \qquad
        \rho_0\sim T_H ,
\end{equation}
up to order-one normalization factors.

The relevant length for low-frequency mode counting is not the spatial proper
length, but the optical, or tortoise, length measured with respect to the
asymptotic time \(t\).  The tortoise coordinate is defined by
\begin{equation}
        dr_*={d\rho\over \rho^2-\rho_0^2}.
\end{equation}
For an arbitrary lower endpoint \(\rho_{\rm IR}>\rho_0\), define
\begin{equation}
        L_{\rm opt}(\rho_{\rm IR};\rho_0)
        =
        \int_{\rho_{\rm IR}}^{O(1)}
        {d\rho\over \rho^2-\rho_0^2}.
\end{equation}
The notation is meant to emphasize that \(\rho_{\rm IR}\) is simply the lower
endpoint of the effective throat description.  It should not yet be identified
with the finite $N$ cutoff.

There are two useful regimes.  If the endpoint lies in the \(AdS_2\)-like
part of the throat, well outside the near-horizon Rindler cap,
\begin{equation}
        \rho_{\rm IR}\gg \rho_0 ,
\end{equation}
then
\begin{align}
        L_{\rm opt}(\rho_{\rm IR};\rho_0)
        &\simeq
        \int_{\rho_{\rm IR}}^{1}{d\rho\over \rho^2}
        \simeq
        {1\over \rho_{\rm IR}} .
\end{align}
On the other hand, if the semiclassical throat is followed down to the
near-horizon scale, say \(\rho_{\rm IR}=c\,\rho_0\) with \(c>1\) fixed, then
\begin{align}
        L_{\rm opt}(\rho_{\rm IR};\rho_0)
        &\sim
        {1\over \rho_0}
        \sim
        {1\over T_H},
\end{align}
up to order-one factors.  If one further resolves the very near-horizon
Rindler region down to a stretched horizon, additional logarithmic factors may
appear, but the power-law enhancement in the near-extremal limit is
\(1/T_H\).  This is the origin of the large density of low-frequency throat
modes.

\paragraph{WKB counting of throat modes}

The one-particle modes in the throat can be estimated by the usual WKB
counting.  Let \(a\) denote a channel label, including the bulk species and a
fixed low-energy set of angular momentum or transverse quantum numbers.  For
each channel, the radial equation can be written schematically in
Schrödinger form,
\begin{equation}
        \left[
        -\partial_{r_*}^2+V_a(\rho)
        \right]\psi_{a,\omega}
        =
        \omega^2\psi_{a,\omega}.
\end{equation}
The corresponding local radial WKB momentum is
\begin{equation}
        k_a(\rho,\omega)
        =
        \sqrt{\omega^2-V_a(\rho)}
\end{equation}
in the classically allowed region.

For fixed \(\omega\), the WKB counting function, namely the number of radial
modes in channel \(a\) with frequency below \(\omega\), is
\begin{equation}
        N_a(\omega)
        \simeq
        {1\over \pi}
        \int_{r_*(\rho_{\rm IR})}^{r_*(O(1))}
        dr_*\, k_a(\rho(r_*),\omega).
\end{equation}
Equivalently,
\begin{equation}
        {dN_a\over d\omega}
        \simeq
        {1\over \pi}
        \int_{r_*(\rho_{\rm IR})}^{r_*(O(1))}
        dr_*\,{\partial k_a(\rho(r_*),\omega)\over\partial\omega}.
\end{equation}
Thus the number of modes below a frequency cutoff \(\Lambda\) is
\begin{align}
        n_a(\omega<\Lambda)
        &=
        N_a(\Lambda)
        =
        \int_0^\Lambda d\omega\,{dN_a\over d\omega}.
\end{align}
This is the precise meaning of the notation \(n_a(\omega<\Lambda)\).  One
should not write an additional integral of \(k_a\) over \(\omega\); the
\(\omega\)-integral acts on the density of states, equivalently on
\(\partial k_a/\partial\omega\).

For the low-energy channels relevant to the parametric estimate, the
effective potential is redshift-suppressed over most of the long throat.
Hence, over the dominant part of the optical interval,
\begin{equation}
        V_a(\rho)\ll \omega^2,
        \qquad
        k_a(\rho,\omega)\simeq \omega,
        \qquad
        {\partial k_a\over\partial\omega}\simeq 1 .
\end{equation}
Therefore
\begin{align}
        {dN_a\over d\omega}
        &\simeq
        {1\over \pi}
        \int_{r_*(\rho_{\rm IR})}^{r_*(O(1))}dr_*
        =
        {L_{\rm opt}(\rho_{\rm IR};\rho_0)\over \pi},
\end{align}
and hence
\begin{equation}
        n_a(\omega<\Lambda)
        \simeq
        {\Lambda\over \pi}
        L_{\rm opt}(\rho_{\rm IR};\rho_0).
\end{equation}
Up to order-one numerical factors,
\begin{equation}
        n_a(\omega<\Lambda)
        \sim
        \Lambda L_{\rm opt}(\rho_{\rm IR};\rho_0).
\end{equation}

If \(K(\Lambda)\) denotes the finite number of species and transverse
channels included below the cutoff \(\Lambda\), then the total number of
low-energy throat modes is
\begin{equation}
        N_{\rm throat}(\omega<\Lambda)
        \sim
        K(\Lambda)\,\Lambda\,L_{\rm opt}(\rho_{\rm IR};\rho_0).
\end{equation}
In the estimate below we keep only a fixed low-energy set of bulk fields and
partial waves, so \(K(\Lambda)\Lambda=O(1)\).  The important scaling is
therefore
\begin{equation}
        N_{\rm throat}(\omega<\Lambda)
        \sim
        L_{\rm opt}(\rho_{\rm IR};\rho_0).
\end{equation}

If the semiclassical throat is continued down to the near-horizon scale
\(\rho_{\rm IR}\sim \rho_0\), this gives
\begin{equation}
        N_{\rm throat}^{\rm semi}
        \sim
        {1\over T_H}.
\end{equation}
This is not a statement about the exact many-body level spacing of the finite
\(N\) CFT.  It is a statement about the number of independent one-particle
modes that local bulk EFT would assign to the near-\(AdS_2\) throat.

\subsection{Thermodynamic consistency of the throat}
\label{subsec:throat-consistency}

The number \(N_{\rm throat}(\omega<\Lambda)\) counts one-particle modes,
not their thermal entropy. To test the fixed-background description, we
compare the energy of thermally populated throat modes with the classical
black-hole energy above extremality. We keep a fixed number \(K=O(1)\) of
light, neutral channels and exclude the final Rindler layer, which is
regulated separately by the ordinary brick wall.

\paragraph{Thermal excitations in the throat}
In the continuum WKB approximation, the free energy at temperature
\(T\) is
\begin{align}
  F_{\rm th}(T)
  &\simeq \frac{K L_{\rm opt}}{\pi}\,
  T\int_0^\Lambda d\omega\,\log\!\left(1-e^{-\omega/T}\right)
  \simeq -\frac{\pi K}{6}L_{\rm opt}T^2,
  \qquad \Lambda\gg T.
\end{align}
Differentiating at fixed background, and only then setting \(T=T_H\), gives
\begin{align}
  E_{\rm th}(T_H)&\simeq \frac{\pi K}{6}L_{\rm opt}T_H^2,
  &
  S_{\rm th}(T_H)&\simeq \frac{\pi K}{3}L_{\rm opt}T_H.
  \label{eq:throat-thermal}
\end{align}
The cutoff \(\Lambda\) drops out because modes with \(\omega\gg T_H\)
are thermally suppressed. For the throat followed down to
\(\rho_{\rm IR}=c\rho_0\), with \(c>1\) fixed, \(L_{\rm opt}\sim T_H^{-1}\)
and hence
\begin{equation}
  E_{\rm th}\sim K T_H,
  \qquad S_{\rm th}\sim K.
\end{equation}
Thus the thermal entropy does not grow as \(1/T_H\), even though the
number of modes below a fixed \(\Lambda\) does.\footnote{
These formulas use the continuum WKB approximation for light bosonic
channels. Effective potentials and discrete levels can change the
numerical coefficients. The parametric estimate concerns the family with
fixed \(\rho_{\rm IR}/\rho_0\); it should not be extrapolated to a
separately capped throat with \(T_H L_{\rm opt}\ll1\), where the discrete
spectrum must be used.
}

\paragraph{Comparison with the near-extremal excitation energy}
Let \(E_0\) and \(S_0\) be the classical extremal energy and entropy at
fixed charge. For an RN-type black hole, the leading classical
near-extremal relations are
\begin{align}
  \Delta S_{\rm BH}^{\rm cl}
  &\equiv S_{\rm BH}^{\rm cl}-S_0
   =\mathcal C T_H+O(T_H^2),\\
  \Delta E_{\rm BH}^{\rm cl}
  &\equiv E_{\rm BH}^{\rm cl}-E_0
   =\frac{\mathcal C}{2}T_H^2+O(T_H^3).
  \label{eq:near-extremal-excitation}
\end{align}
Here \(\mathcal C>0\) is the fixed-charge entropy slope at extremality;
the second relation follows from the first law. With the macroscopic
length scales fixed as above, \(\mathcal C\sim S_0\sim S_{\rm BH}\).
These relations follow directly from the higher-dimensional black-hole
thermodynamics, without using a JT description
\cite{Preskill:1991tb,Almheiri:2016fws}.

For the throat excitations to be a small perturbation of the same
near-extremal background, their energy must be small compared with
\(\Delta E_{\rm BH}^{\rm cl}\), not merely with the total black-hole mass:
\begin{equation}
  \frac{E_{\rm th}}{\Delta E_{\rm BH}^{\rm cl}}
  \simeq \frac{\pi K L_{\rm opt}}{3\mathcal C}\ll1.
  \label{eq:throat-consistency}
\end{equation}
Indeed, adding an energy \(\delta E\) at fixed charge changes the
classical temperature by
\begin{equation}
  \frac{\delta T_H}{T_H}
  \simeq \frac{\delta E}{\mathcal C T_H^2}.
\end{equation}
Thus an energy small compared with the total mass can nevertheless
change the near-extremal temperature and the deep throat by order one.
In this thermal gas description, the condition for parametric control is
\begin{equation}
  K L_{\rm opt}\ll\mathcal C,
\end{equation}
and the approximation reaches its limit at
\begin{equation}
  L_{\rm opt}\sim\frac{\mathcal C}{K}\sim S_{\rm BH}
  \qquad (K=O(1)).
  \label{eq:throat-breakdown-length}
\end{equation}
This is a backreaction consistency estimate for thermal excitations on a
fixed classical background, not a comparison of one-particle mode number
with \(e^{S_{\rm BH}}\).\footnote{
Here \(E_{\rm th}\) is the thermal excitation energy in the
fixed-background gas approximation, not a separately added contribution
to the black-hole mass. A complete calculation of the corrected energy
also includes vacuum polarization and gravitational backreaction.
Although \eqref{eq:throat-thermal} and
\eqref{eq:near-extremal-excitation} give the same parametric ratio
\(S_{\rm th}/\Delta S_{\rm BH}^{\rm cl}\), we do not treat
\(\Delta S_{\rm BH}^{\rm cl}\) as an independent Hilbert-space capacity.
}

For \(\rho_{\rm IR}=c\rho_0\), the enhancement
\(L_{\rm opt}\sim T_H^{-1}\) therefore gives the temperature scale
\begin{equation}
  T_H\sim\frac{K}{\mathcal C}\sim S_{\rm BH}^{-1}
  \sim N_{\rm eff}^{-2}
  \qquad (K=O(1)).
  \label{eq:throat-breakdown-temperature}
\end{equation}
Above this scale the thermal throat energy is a parametrically small
correction to the classical excitation energy. At this scale it is not:
\(E_{\rm th}\sim K T_H\), whereas
\(\Delta E_{\rm BH}^{\rm cl}\sim\mathcal C T_H^2\) vanishes faster as
the temperature is lowered. This is why the small thermal entropy
\(S_{\rm th}=O(K)\) does not by itself guarantee the validity of the
fixed-background description.

\paragraph{Relation to the ordinary brick wall}
The ordinary brick wall regulates the final Rindler region. In terms of
the proper distance \(x\) from the horizon,
\begin{equation}
  ds^2\simeq-\kappa^2x^2dt^2+dx^2+r_h^2d\Omega_{d-1}^2,
  \qquad T_{\rm loc}\simeq\frac{1}{2\pi x}.
\end{equation}
Including the full transverse phase space gives
\begin{equation}
  S_{\rm BW}\sim A\int_\epsilon dx\,T_{\rm loc}^{\,d}
  \sim\frac{A}{\epsilon^{d-1}}.
\end{equation}
Matching its scaling to \(S_{\rm BH}\sim A/G_N\) gives
\(\epsilon\sim l_p\). This is the usual local UV cutoff near the horizon.

The throat estimate above instead uses a fixed set of light channels
outside that Rindler layer. It does not sum all transverse modes below
a fixed asymptotic cutoff \(\Lambda\). For the thermal modes at
\(\rho=c\rho_0\),
\begin{equation}
  T_{\rm loc}
  =\frac{T_H}{L_2\rho_0\sqrt{c^2-1}}
\end{equation}
is independent of \(T_H\) at fixed \(c>1\), since
\(\rho_0\propto T_H\). Thus the thermal transverse phase space does not
acquire an additional inverse power of \(T_H\) in this region. The
consistency condition concerns the energy carried by these throat modes
relative to the shrinking near-extremal excitation energy, rather than
the UV entropy divergence in the final Rindler layer.

\paragraph{Interpretation and relation to near-\(AdS_2\) physics}
The scale \eqref{eq:throat-breakdown-temperature} is the familiar
thermodynamic scale at which a typical Hawking quantum has an energy
comparable to the classical energy above extremality
\cite{Preskill:1991tb}. It is parametrically consistent with the
near-\(AdS_2\) backreaction scale \cite{Almheiri:2016fws}; quantum
corrections to near-extremal thermodynamics become important at this
scale \cite{Iliesiu:2020qvm}.

The conclusion is that a fixed classical throat with independently
quantized thermal excitations is no longer parametrically controlled at
\(L_{\rm opt}\sim S_{\rm BH}\), for the family and light channels
considered here. This does not establish a physical wall at a universal
radial position, a strict upper bound on the optical length of a fully
quantum geometry, or a lower bound on its temperature. It identifies the
scale at which the classical throat and its excitations must be treated
together; how the finite-\(N\) theory modifies the deep throat is not
determined by this estimate.

\subsection{Regular monopole cores as dynamical brick walls}
\label{subsec:monopole-brick-wall}

It is useful to point out that the brick-wall boundary condition need not be
viewed only as an artificial prescription.  In suitable matter--gravity
systems, the would-be horizon can be replaced by a smooth classical core,
while the exterior remains very close to the corresponding charged black hole
geometry.  Gravitating magnetic monopoles provide a concrete example of this
phenomenon.
We use these horizonless configurations and their near-core excitations
as semiclassical models of a near-extremal black hole for coarse-grained
exterior probes. They are not themselves an explicit microscopic
finite-\(N\) construction.

Consider a sector with fixed magnetic charge \(Q\) and ADM energy \(E\).
In pure Einstein--Maxwell theory, the Reissner--Nordstr\"om family has an
extremal solution at \(E=|Q|\), subextremal black holes for \(E>|Q|\), and
superextremal naked singular solutions for \(E<|Q|\), in appropriate units.
In an Einstein--Yang--Mills--Higgs theory, however, the singular interior of a
would-be Reissner--Nordstr\"om (RN) solution may be replaced by non-Abelian gauge and Higgs fields.
Then one can have globally regular, horizonless monopole solutions whose
exterior is close to the RN geometry down to a radius just outside the
would-be horizon.

Such solutions were studied in \cite{Breitenlohner:1991aa} in the
gravitating 't Hooft--Polyakov monopole system.  In one branch, regular
monopole solutions approach the extremal RN geometry at the critical point:
outside the monopole core the metric is essentially RN-like, while the core
replaces the would-be singular or horizon region.  There are also additional
branches of solutions, including branches which can extend into a range where
the exterior parameters would correspond, in pure Einstein--Maxwell theory,
to a subextremal RN black hole.  In these solutions the geometry can remain
regular and horizonless because the monopole core lies outside the would-be
horizon.  Some of these branches are expected to be unstable, and we will not
use them as stable microstate geometries.  The point is instead that they
provide explicit classical examples in which a black hole-like exterior is
capped off smoothly by matter fields.

Near the would-be extremal RN geometry, the exterior throat is locally
approximated by
\begin{equation}
        ds^2
        \simeq
        -{\rho^2-\delta^2\over M^2}\,dt^2
        +{M^2\over \rho^2-\delta^2}\,d\rho^2
        +M^2 d\Omega_2^2 .
\end{equation}
For the corresponding RN black hole, the outer horizon would be at
\begin{equation}
        \rho=\delta ,
\end{equation}
and the Hawking temperature would be
\begin{equation}
        T_H={\delta\over 2\pi M^2}.
\end{equation}
In the extremal limit \(\delta\to0\), the horizon scale moves to
\(\rho=0\).  In the regular monopole solution there is no horizon; instead,
the RN approximation breaks down at a core radius outside the would-be
horizon, where the non-Abelian gauge and Higgs fields become important.

For the purposes of the present discussion, one may place the system in
asymptotically AdS spacetime.  This should be viewed mainly as an infrared
regularization: it removes the continuous spectrum associated with an
asymptotically flat region and allows us to discuss normal modes in a compact
setting.  We take the AdS scale and the charge radius \(M\) to be of order
one.  A useful regime is
\begin{equation}
        S_{\rm BH}^{-1}\ll \delta \ll 1,
\end{equation}
or equivalently \(N_{\rm eff}^{-2}\ll \delta\ll 1\), where the exterior is
near extremal but remains above the thermodynamic consistency scale
discussed in subsection~\ref{subsec:throat-consistency}. This lets us
focus on the ordinary brick-wall-like physics associated with the
smooth monopole core.

The semiclassical description around such a regular solution is obtained by
linearizing the fields around the classical background and quantizing the
normal modes.  Since the exterior geometry is close to RN down to the core,
the radial wave equation outside the core is essentially the same as in the
RN exterior.  The only difference from the artificial brick-wall model is that
the inner boundary condition is not imposed by hand.  It is determined by
regularity and by matching to the monopole core.  At the level of leading WKB
mode counting, this replacement changes only order-one details such as the
reflection phase.  The density of redshifted low Killing-energy modes is the
same as in a brick-wall model whose wall is placed at the radius where the RN
approximation breaks down.

Thus, if the monopole core lies at a microscopic proper distance outside the
would-be horizon, the usual brick-wall estimate is recovered.  The large
redshift between the core and infinity produces many modes with small
asymptotic energy but large local energy near the core.  Exciting these modes
gives a semiclassical entropy with the same area scaling as the brick-wall
entropy,
\begin{equation}
        S_{\rm wall}\sim {A\over G_N},
\end{equation}
up to order-one coefficients depending on the species, the precise matching
condition at the core, and the cutoff prescription.  This is the sense in
which the regular monopole core realizes brick-wall physics dynamically.

It is important, however, to separate two notions of entropy.  The regular
monopole solution is a classical background, and a given branch of classical
solutions does not by itself provide an exponentially large set of distinct
geometries.  The entropy just described is instead the semiclassical entropy
of quantum states obtained by exciting modes around the same classical
background.  This is analogous to the brick-wall calculation: the entropy
comes from field-theoretic excitations supported by the redshifted exterior
region, not from an explicitly enumerated set of distinct classical solutions.

From a quantum-mechanical viewpoint, this is natural.  A classical solution
does not correspond to a unique exact quantum state.  Even for a harmonic
oscillator, a classical trajectory is represented semiclassically by a
coherent state, and small quantum excitations around it need not be
distinguishable by coarse exterior probes.  Similarly, for a regular monopole
background close to a would-be RN black hole, low-energy excitations localized
in the highly redshifted region near the core can be invisible to observers
far outside the would-be horizon, while still contributing to the number of
states associated with that semiclassical exterior.

If the solution were the unique ground state at fixed charge, this
degeneracy might be absent or strongly reduced.  But the near-extremal
situation is different: \(E\) is above the minimal energy allowed at fixed
\(Q\), and the redshifted exterior modes can be excited without appreciably
changing the geometry seen far from the core.  The resulting state counting
is precisely the semiclassical brick-wall counting, now realized with a
regular matter core rather than an artificial boundary.

This also clarifies the relation to fuzzball ideas.  A regular monopole core
of this type is not a complete fuzzball ensemble: one does not obtain an
exponentially large family of distinct smooth classical geometries with the
same asymptotic charges.  Nevertheless, it is fuzzball-like in a weaker
semiclassical sense.  The would-be horizon is absent, the exterior is
black hole-like, and the cap replaces the artificial brick-wall boundary by a
regular classical core.  The brick-wall entropy is then interpreted as the
entropy of semiclassical excitations supported outside this cap.

\section{Conclusions}

We have argued that the brick-wall model should be viewed not as a literal
hard surface at the horizon, but as a simple effective parametrization of
the finite $N$ departure from the semiclassical near-horizon continuum.
In this interpretation the wall regulates the infinite density of
near-horizon bulk EFT modes and replaces it by a large, but finite,
stretched-horizon reservoir.

A free Dirichlet wall is of course reflecting.  The main point of this
paper is that this fact should be separated from the inclusive response
seen by exterior simple probes.  Once the near-wall modes are treated as
an interacting reservoir, an incoming coherent one-particle excitation can
be converted into many complicated near-wall degrees of freedom.  The
exterior simple sector can then see an absorptive, black hole-like
response, even though the microscopic evolution remains unitary.

The same picture gives a simple effective account of Hawking radiation.
For the Hartle--Hawking-like state, the reservoir and the exterior modes
are in thermal equilibrium and there is no net luminosity.  For the
Unruh-like state, the reservoir is thermally or typically populated while
the ordinary exterior incoming modes are in their vacuum.  The system then
radiates as an ordinary hot object, with the emitted quanta filtered by
the usual greybody potential.

At finite \(N\), however, the near-horizon sector cannot be an ideal
continuum absorber at arbitrarily fine resolution.  Therefore perfect
absorption is at best a coarse-grained or early-time approximation.  In
the brick-wall description the failure of perfect absorption appears as
reflection or echoes; in a more microscopic finite $N$ description the
return may be scrambled and incoherent, but it is a response of the
stretched-horizon degrees of freedom on the scrambling time scale
\(\kappa^{-1}\log S_{\rm BH}\), not a Heisenberg-time recurrence.  The
size and detailed waveform of the return signal are model-dependent, but
we expect the effect to be of order \(N^0\), not
\(e^{-S_{\rm BH}}\)-suppressed.

Finally, we discussed how brick-wall-like behavior may arise dynamically
rather than by imposing a boundary condition by hand.  Regular matter
cores, such as gravitating monopole cores, can cap off a black hole-like
exterior and provide an effective inner boundary.  This is not a full
fuzzball construction, but it illustrates the weaker mechanism emphasized
here: the smooth horizon of the continuum EFT can be replaced by finite
microscopic structure while preserving a black hole-like exterior response
over the appropriate coarse-grained time scales.


\section*{Acknowledgements}

The author would like to thank S. Chikazawa, T. Kawamoto, Y. Nakayama and S. Sugishita for their useful discussions.
This work was supported by JSPS KAKENHI Grant Number 	24K07048.

\hspace{1cm}

\bibliographystyle{utphys}
\bibliography{main202505.bib}

\end{document}